\documentclass[prb,twocolumn,10pt,aps,nofootinbib,superscriptaddress]{revtex4-1}
 \usepackage{verbatim}
 \usepackage{amsmath}
 \usepackage{amssymb}
 \usepackage{amsthm}
 \usepackage{dsfont}
 \usepackage{latexsym}
 \usepackage{amsfonts}
 \usepackage{epsfig}
 \usepackage{xcolor}
 \definecolor{darkblue}{rgb}{0,0,.5}
 \usepackage[linktocpage, colorlinks=true, linkcolor=darkblue, citecolor=darkblue]{hyperref}
 \usepackage[all]{hypcap}

\usepackage[T1]{fontenc}
\usepackage[english]{babel}
\usepackage{graphicx}
\usepackage{array}

\usepackage{ragged2e}
\usepackage{fancyhdr}
\usepackage{booktabs}
\usepackage{url}
\usepackage{dsfont}
\usepackage{tcolorbox}
\usepackage{import}
\usepackage{soul}

\usepackage{media9}
\usepackage{tikz}
\usepackage[overlay]{textpos}
\usepackage[makeroom]{cancel}
\usepackage{framed,color,xcolor}
\usepackage{pstricks}
\usepackage[colorinlistoftodos]{todonotes}
\usepackage{ulem}
\usepackage{url}
\usepackage{hyperref}
\usepackage{placeins}

\DeclareMathSizes{9}{9}{6.5}{6.5}

\definecolor{turquoise3}{rgb}{0,0.77,0.8}
\definecolor{myred}{RGB}{220,1,50}
\definecolor{mygreen}{RGB}{0,150,100}

\definecolor{dmet_turq}{RGB}{6,205,214}
\definecolor{dmet_lila}{RGB}{86,59,204}
\definecolor{dmet_pink}{RGB}{242,0,95}

\newcommand{\bra}[1]{\langle #1|}
\newcommand{\ket}[1]{|#1\rangle}

\begin{document}
	\title{Density-matrix embedding theory study of the one-dimensional Hubbard-Holstein model}
	
	\author{Teresa E.~Reinhard}
	\email{teresa.reinhard@mpsd.mpg.de}
	\affiliation{Max Planck Institute for the Structure and Dynamics of Matter, Center for Free Electron Laser Science, 22761 Hamburg, Germany}
	\author{Uliana Mordovina}
	\affiliation{Max Planck Institute for the Structure and Dynamics of Matter, Center for Free Electron Laser Science, 22761 Hamburg, Germany}
	\author{Claudius Hubig}
	\affiliation{Max Planck Institute of Quantum Optics, Hans-Kopfermann-Str. 1, 85748 Garching, Germany}
	\author{Joshua S.~Kretchmer}
	\affiliation{Division of Chemistry and Chemical Engineering, California Institute of Technology, Pasadena, CA 91125}
	\author{Ulrich Schollw\"ock}
	\affiliation{Department of Physics and Arnold Sommerfeld Center for Theoretical Physics,
		Ludwig-Maximilians-Universit\"at M\"unchen, Theresienstrasse 37, 80333 M\"unchen, Germany}
	\author{Heiko Appel}
	\affiliation{Max Planck Institute for the Structure and Dynamics of Matter, Center for Free Electron Laser Science, 22761 Hamburg, Germany}
	\author{Michael A.~Sentef}
	\affiliation{Max Planck Institute for the Structure and Dynamics of Matter, Center for Free Electron Laser Science, 22761 Hamburg, Germany}
	\author{Angel Rubio}
	\affiliation{Max Planck Institute for the Structure and Dynamics of Matter, Center for Free Electron Laser Science, 22761 Hamburg, Germany}
	\affiliation{Center for Computational Quantum Physics (CCQ), Flatiron Institute, 162 Fifth Avenue, New York NY 10010}

	\date{\today}
\begin{abstract}
We present a density-matrix embedding theory (DMET) study of the one-dimensional Hubbard-Holstein model, which is paradigmatic for the interplay of electron-electron and electron-phonon interactions. Analyzing the single-particle excitation gap, we find a direct Peierls insulator to Mott insulator phase transition in the adiabatic regime of slow phonons in contrast to a rather large intervening metallic phase in the anti-adiabatic regime of fast phonons. We benchmark the DMET results for both on-site energies and excitation gaps against density-matrix renormalization group (DMRG) results and find excellent agreement of the resulting phase boundaries. We also compare the fully quantum treatment of phonons against the standard Born-Oppenheimer (BO) approximation. The BO approximation gives qualitatively similar results to DMET in the adiabatic regime, but fails entirely in the anti-adiabatic regime, where BO predicts a sharp direct transition from Mott to Peierls insulator, whereas DMET correctly shows a large intervening metallic phase. This highlights the importance of quantum fluctuations in the phononic degrees of freedom for metallicity in the one-dimensional Hubbard-Holstein model. 
\end{abstract}

\maketitle

\section{Introduction}\label{section:intro}
The interplay of competing interactions is a central theme of quantum many-body physics. In quantum materials, the electron-electron (el-el) and electron-phonon (el-ph) interactions naturally compete against each other. This is most easily understood by noting that el-el Coulomb repulsion is generically repulsive, whereas el-ph interactions can lead to effectively attractive el-el interactions, as highlighted by the Cooper pairing mechanism in conventional superconductors \cite{bardeen_theory_1957}. In strongly correlated low-dimensional materials, the competition between el-el and el-ph interactions has led to longstanding debates, such as the origin of high-temperature superconductivity and the anomalous normal states observed in entire classes of materials \cite{shen_missing_2004,lanzara_evidence_2001}.

At the same time, competing interactions lead to competing groundstates and phase transitions, as exemplified by the complex phase diagrams in correlated oxides \cite{keimer_quantum_2015}.
This competition and sensitivity to parameter changes poses a major roadblock on the way towards reliable numerical solutions for the quantum many-body electron-phonon problem. The simplest generic el-ph model is the Hubbard-Holstein Hamiltonian \cite{fradkin1983phase, von1995hubbard}. Advanced numerical methods for its solution have been developed over the past decades that are accurate in certain regimes but cannot be easily applied in other cases. Among these are Quantum Monte Carlo (QMC) \cite{clay_intermediate_2005, hardikar_phase_2007, hohenadler_excitation_2013, weber_two-dimensional_2018, nowadnick_competition_2012,johnston_determinant_2013}
schemes, the density-matrix renormalization group (DMRG) \cite{fehske_metallicity_2008}, as well as dynamical mean-field theory (DMFT) \cite{werner_efficient_2007, murakami_ordered_2013,bauer_competition_2010}

A new, promising method that has recently been added to this arsenal is density-matrix embedding theory (DMET) \cite{knizia_density_2012, knizia_density_2013}, which in some sense bridges DMRG and DMFT-related methods. DMET has the advantage of being numerically less demanding than DMRG and at the same time being a good descriptor for one-dimensional systems (opposed to DMFT). DMET has been benchmarked against other methods for the 2D Hubbard model \cite{zheng_ground-state_2016,leblanc_solutions_2015}
and recently a systematic extension of DMET towards DMFT was proposed \cite{fertitta_rigorous_2018}. However, for the Hubbard-Holstein model only one DMET study has been published to the best of our knowledge so far, which compared ground state energies for the 1D Hubbard-Holstein model against DMRG results \cite{sandhoefer_density_2016}.

In this work, we perform extensive comparisons of DMET against both DMRG and Born-Oppenheimer (BO) results for the 1D Hubbard-Holstein model. In particular, we extend the previous comparisons to excitation gaps and the difference of the electron density between neighbouring sites, which indicates the existence of a charge-density wave (CDW). This allows us to construct DMET phase diagrams that are compared directly against DMRG. 

The paper is organized as follows: In section \ref{section:model} we briefly introduce the Hubbard-Holstein model. In section \ref{section:DMET_el} we summarize the DMET method as originally constructed for purely electronic systems. In section \ref{section:DMET_phon} we explain the extension of the DMET method for electron-phonon systems. In section \ref{section:obs} we discuss the phases of the Hubbard-Holstein model and possible observables to determine these phases. In section \ref{section:results}, we discuss our DMET results and benchmark them against a DMRG calculation. Moreover we then benchmark a semiclassical Born-Oppenheimer calculation against DMET and show when the Born-Oppenheimer approximation fails.

\section{Hubbard-Holstein model}\label{section:model}
The Hubbard-Holstein model is described by the Hamiltonian
\begin{align}\label{formula:hub_hol}
\hat{H}_{\rm el-ph} &= \underbrace{\sum_{<i,j>, \sigma} t \hat{c}_{i\sigma}^{\dagger}\hat{c}_{j\sigma}}_{A}
+\underbrace{\sum_{i} U \hat{n}_{i \uparrow} \hat{n}_{i \downarrow}}_{B} \nonumber\\
&+ \underbrace{\sum_i \omega_0 \hat{a}_i^{\dagger} \hat{a}_i}_{C}
+\underbrace{\sum_{i,\sigma} g \hat{n}_{i,\sigma}(\hat{a}_i^{\dagger}+\hat{a}_i)}_{D}.
\end{align}
Here, $\hat{c}_{i\sigma}^{\left(\dagger\right)}$ is the electronic particle creation (annihilation) operator on site $i$, where $\sigma$ ( $=$ $\uparrow$, $\downarrow$) is the spin degree of freedom, $\hat{n}_{i \uparrow\left(\downarrow\right)} = \hat{c}_{i\uparrow\left(\downarrow\right)}^{\dagger}\hat{c}_{i\uparrow\left(\downarrow\right)}$ the spin-up (spin-down) particle number operator and $\hat{a}_i^{\left(\dagger\right)}$ is the phononic particle creation (annihilation) operator.\\
The kinetic energy of the electrons is approximated as a simple next-neighbor hopping term (A) and the electron-electron interaction is assumed to be a purely local Hubbard $U$ term (B). The phonons are approximated by harmonic oscillators (C) which are bilinearly coupled to the density of the electrons (D). One phonon mode is considered per electronic site.

\section{DMET for electronic systems}\label{section:DMET_el}
Generally, when trying to solve the Schr\"odinger equation 
\begin{align}\label{formula:schroed_el}
\hat{H}_{\rm el}\ket{\Psi} &= E\ket{\Psi},\\
\hat{H}_{\rm el} &= \hat{T}_{\rm el} + \hat{U}_{\rm el}
\end{align}
for a given general electronic Hamiltonian, the well-known problem of the exponential wall of quantum mechanics occurs, making the costs of the calculation grow exponentially with system size. Even though there are wave function methods that scale this problem down to polynomial growth, it is still a fact that normal wave function methods grow fast with the size of the regarded system, making it hard to describe large systems. One possible way to circumvent this problem is the embedding idea that is used in different methods including DMET: Instead of solving the Schr\"odinger equation for the whole system, a small subsystem is chosen, which is small enough to be solved efficiently. The idea of DMET is then, to include the interactions of the rest of the system with this subsystem, which will from now on be called ``impurity'', in the embedding step. This is equivalent to a complete active space calculation in quantum chemistry, assuming that the impurity part of the system is in the active space. This way, we divide the system into two unentangled parts: The so-called embedded system, consisting of the impurity and those parts of the rest of the system interacting with the impurity, and the environment that consists of those parts of the system that do not interact with the impurity directly. Then, by solving the embedded system, the physics of the impurity, including interactions with the rest of the system (and with that, also finite size effects and the influence of the boundaries) can be computed effectively with an accurate wave function method, since the embedded system is typically much smaller than the originally considered system. In this work, we target lattice Hamiltonians. Thus, the following derivation will be shown for the lattice site basis.\\
Mathematically, the projection from a lattice basis to the impurity plus active space basis can be formulated with the help of the Schmidt decomposition:
\begin{align}\label{formula:schmidt_decomp}
\ket{\Psi}&=\sum_{a}^{ 4^{ N_{ \rm imp}}} \sum_{b}^{4^{N - N_{\rm imp }}} \Psi_{ab} \ket{\mathcal{\tilde{A}}_a}\ket{\tilde{\mathcal{B}}_b}\\
&= \sum_{a}^{ 4^{N_{\rm imp}}} \sum_{b}^{4^{N - N_{\rm imp }}} \sum_{i}^{ 4^{N_{\rm imp}}}U_{ai}\lambda_i V_{ib}\ket{\mathcal{\tilde{A}}_a}\ket{\tilde{\mathcal{B}}_b}\\
&= \sum_{i}^{4^{N_{\rm imp}}}\lambda_i \ket{\mathcal{A}_i} \ket{\mathcal{B}_i}
\end{align}
Every wave function can be decomposed into two parts $\ket{\mathcal{\tilde{A}}_a}$ and $\ket{\tilde{\mathcal{B}}_b}$ where the former are defined on the impurity and the latter on the rest of the system. Both parts are coupled to each other by the transition matrix $\Psi_{ab}$. Performing a singular value decomposition of this matrix leads then to a new basis consisting of the many-body states $\ket{\mathcal{A}_i}$ and $\ket{\mathcal{B}_i}$. The number of many-body states describing the whole system is then $4^{2\cdot N_{\rm imp}}$, where $N_{\rm imp}$ is the number of lattice sites describing the impurity.\\
Knowing the many-body states $\ket{\mathcal{A}_i}$ and $\ket{\mathcal{B}_i}$, a projection 
\begin{align}\label{formula:proj_el}
P_{\rm el} = \sum_{ij}\ket{\mathcal{A}_i}\ket{\mathcal{B}_i}\bra{\mathcal{B}_j}\bra{\mathcal{A}_j}
\end{align}
can be defined, that projects the Hamiltonian onto a new basis 
\begin{align}\label{formula:h_emb_el}
\hat{H}^{\rm emb}_{\rm el} = P_{\rm el}^{\dagger}\hat{H}_{\rm el}P_{\rm el}.
\end{align}
$\hat{H}^{\rm emb}_{\rm el}$ is then a many-body Hamiltonian of dimension $4^{2\cdot N_{\rm imp}}$ and thus can be diagonalized efficiently with a chosen wave function method.\\
Unfortunately though, in order to find the active space states $\ket{\mathcal{B}_i}$, the whole transition matrix $\Psi_{ij}$, that is, the whole wave function $\ket{\Psi}$ needs to be known. This is why, instead of using the exact projection, we have to approximate it in order to find the embedding Hamiltonian. Note that we only approximate the projection of the Hamiltonian into a new basis and not the Hamiltonian itself.\\
We find the projection by choosing a system that can be solved even for big system sizes, namely the kinetic part of the whole Hamiltonian of Eq.~(\ref{formula:schroed_el}) $\hat{T}_{\rm el}$, whose ground state wave function is a Slater determinant.
It can be shown that a Slater determinant can further be decomposed to \cite{knizia_density_2013}
\begin{align}\label{formula:slater}
	\ket{\Phi}=\underbrace{\sum_{i,l}^{N_{\rm imp}}\Phi_{il}\ket{A_i}\ket{B_l}}_{\rm emb}+
    \underbrace{\sum_{j}^{N-2 N_{\rm imp}}\phi_j\ket{\tilde{B}_j}}_{\rm environment}.
\end{align}
Here, the $\ket{A_i}$ are many body states defined on the first $N_{\rm imp}$ impurity sites and the $\ket{B_l}$ are many body states defined on the remaining $N-N_{\rm imp}$ lattice sites. The environment, that is, the $\ket{\tilde{B}_j}$, are many body states defined on the whole system, i.e. on $N$ lattice sites, where none of the particles is on the impurity lattice sites. Thus, in order to describe the physics on the impurity, only the first part of the wave function, which consists of $2N_{\rm imp}$ many body states, is needed.\\
From this Slater determinant $\ket{\Phi}$ and neglecting the mixing terms, we define the projection as
\begin{align}\label{formula:proj_el_slater}
	P_{\rm el}
	= \sum_{i,l}^{N_{\rm imp}}\ket{A_i}\ket{B_l}\bra{B_{l}}\bra{A_{i}} +  \sum_{j}^{N-2N_{\rm imp}}\ket{\tilde{B}_j} \bra{\tilde{B}_{j}},
\end{align}
yielding the embedding Hamiltonian for the interacting system
\begin{align}\label{formula:h_emb_el_2}
	P_{\rm el}^{\dagger}\hat{H}_{\rm el}P_{\rm el} =& \sum_{i,l,i',l'}\ket{A_i}\ket{B_l}\bra{B_l}\bra{A_i}\hat{H}_{\rm el}\ket{A_{i'}}\ket{B_{l'}}\bra{B_{l'}}\bra{A_{i'}}\nonumber\\
    + &\sum_{jj'}\ket{\tilde{B}_j}\bra{\tilde{B}_j}\hat{H}_{\rm el}\ket{\tilde{B}_{j'}}\bra{\tilde{B}_{j'}}\nonumber\\ 
    +&\sum_{i,l,j'}\ket{A_i}\ket{B_l}\bra{B_l}\bra{A_i}\hat{H}_{\rm el}\ket{B_{j'}}\bra{B_{j'}} + h.c.\nonumber\\
    =& \hat{H}^{\rm emb}_{\rm el} + \hat{H}^{\rm env}_{\rm el} + \hat{H}^{\rm emb-env}_{\rm el} ,
\end{align}
where $\hat{H}_{\rm el}^{\rm emb}$ is the Hamiltonian describing the impurity plus its interaction with the bath. $\hat{H}_{\rm el}^{\rm env}$ describes the environment which just yields an energy shift due to the unentangled rest of the environment system. The last term, $\hat{H}_{\rm el}^{\rm emb-env}$ will also be neglected as it would vanish if the projection was exact.
\\
Put differently, we approximate the active space belonging to the chosen impurity for the interacting with the active space of the non-interacting system. In order to improve the projection, we find a one body potential $\hat{V}_{\rm el}$ that we add to the kinetic part of the Hamiltonian
\begin{align}\label{formula:h_proj_el}
\hat{H}^{\rm proj}_{\rm el} = \hat{T}_{\rm el} + \hat{V}_{\rm el}
\end{align}
making the active space of the interacting and the non-interacting system as similar as possible. We find $\hat{V}_{\rm el}$ by minimizing the distance between the reduced one-particle density matrices of the interacting and the non-interacting systems on the impurity space. That is, we minimize the following expression:
\begin{equation}\label{formula:min_el}
\text{min} \left| \sum_{i,j\in \mathrm{emb}} \left< \Psi_{\rm emb}|c_i^{\dagger}c_j|\Psi_{\rm emb} \right> - \left<\Phi|c_i^{\dagger}c_j|\Phi\right> \right|,
\end{equation}
where $\ket{\Psi_{\rm emb}}$ is the ground state wave function of $\hat{H}_{\rm emb}$ and $\ket{\Phi}$ is the ground state wave function of $\hat{H}_{\rm proj}$.

\section{DMET for coupled phonon-electron problems}\label{section:DMET_phon}
In order to generalize the procedure described in section \ref{section:DMET_el}, we follow the derivation presented in \cite{sandhoefer_density_2016}. We need to find a projection $P$ that projects the coupled electron-phonon Hamiltonian
\begin{align}\label{formula:h_el_ph_2}
\hat{H}_{\rm el-ph}=\hat{T}_{\rm el} + \hat{U}_{\rm el} + \hat{T}_{\rm ph} + \hat{U}_{\rm el-ph}.
\end{align}
onto an embedding basis,
\begin{align}\label{formula:h_emb_el_ph}
\hat{H}^{\rm emb}_{\rm el-ph} = P^{\dagger}\hat{H}_{\rm el-ph}P.
\end{align}
\begin{figure}[t]
	\includegraphics[width=.5\textwidth]{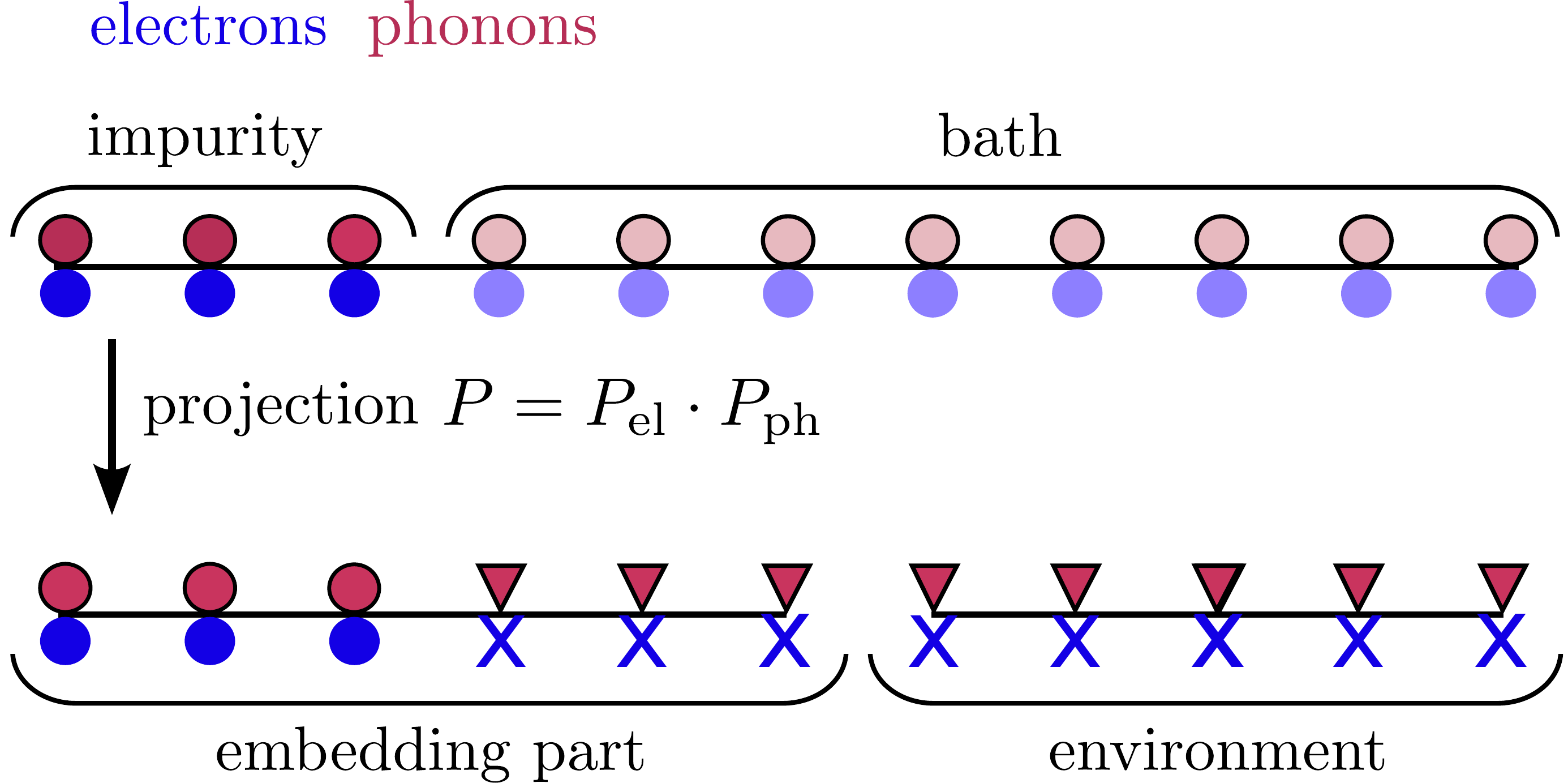}
    \caption{Visualization of the decomposition of the electron-phonon system via the projection $P$: Starting with a 1D lattice in real space that on each site has both electronic (blue) as well as phononic (red) degrees of freedom, we choose one part of the system that is from then on called impurity, whereas the rest is the environment. The electronic and the phononic sites are treated equally in this scheme by defining the full projection as a product of one projection for the electrons and one for the phonons. This projection then leaves the basis on the (electronic and phononic) impurities the same (a real space lattice), whereas it projects the (electronic and phononic) environment degrees of freedom into a new basis set whose physical meaning is abstract. Within this new basis set, the environmental degrees of freedom can be divided into those interacting with the impurity and those not interacting with the impurity, called environment. The system containing the impurity and the basis sites interacting with the impurity is called embedded system. In our calculation, we discard the environment system and only calculate the embedded system.}
    \label{pic:basis_and_projection}
\end{figure}
\\
Similar to the purely electronic case, we find this projection by assuming a non-interacting system, which then allows us to make a product ansatz between the electronic and phononic degrees of freedom. This procedure is visualized in Figure \ref{pic:basis_and_projection}.\\
Instead of finding one projection for the coupled electron-phonon system, we thus have to find two projections, one for the electrons and one for the phonons.
The electronic projection is then, as in the original scheme, approximated by the ground state of $\hat{H}^{\rm proj}_{\rm el}$ defined in Eq.~(\ref{formula:h_proj_el}).\\
In order to find the projection for the phononic degrees of freedom we consider a shifted harmonic oscillator, which is described by a Hamiltonian of the same form as the electron-phonon Hamiltonian of Eq.~(\ref{formula:hub_hol}):
\begin{align}\label{formula:h_ph}
\hat{H}_{\rm ph} = \sum_i \omega_0 \hat{a}_i^{\dagger} \hat{a}_i + \sum_{i,\sigma} g_i (\hat{a}_i^{\dagger}+\hat{a}_i) = \hat{T}_{\rm ph} + \hat{C}_{\rm ph}
\end{align}
The ground state wave function of this Hamiltonian is the product state of coherent states on each lattice site $k$:
\begin{align}\label{formula:coh_state}
	\ket{Z} &= \bigotimes_k \alpha_k\ket{z_k}\\
	\ket{z_k} &= e^{a_k^{\dagger}z_k}\ket{0} = e^{-|z_k|^2/2}\sum_{j=0}^{\infty} \frac{(z_k)^j}{\sqrt{j!}}\ket{j}, 
\end{align}
where $z_k = \left<\hat{a}_k^{\dagger}+\hat{a_k}\right>$ is the shift of the phonon mode from the initial position on the lattice site $k$.
Note that the state $\ket{z_k}$ is a superposition of all possible phononic Fock numbers states on site $k$.
In the original Hamiltonian $\hat{H}_{\rm el-ph}$, defined in Eq.~(\ref{formula:hub_hol}), due to the coupling term between electrons and phonons, the total number of phonons is not conserved (which makes sense as they are only quasi-particles describing the lattice vibrations of the solid).
As the coherent state defined here in Eq.~(\ref{formula:coh_state}) also does not obey phononic particle number conservation, it is well suited to describe our problem.
\\
Similar to the electronic case, this approximate Hamiltonian yields the phononic projection as:
\begin{align}\label{formula:proj_ph}
P_{\rm ph} =& \sum_{i,l}^{N_{\rm imp}} \ket{A_i^{\rm ph}}\ket{B_l^{\rm ph}}\bra{B_l^{\rm ph}}\bra{A_i^{\rm ph}}\nonumber\\
&+ \sum_{j}^{N- 2N_{\rm imp}}\ket{\tilde{B}_j^{\rm ph}}\bra{\tilde{B}_j^{\rm ph}}.
\end{align}
Knowing the two projections, we are now able to find the embedding Hamiltonian of the coupled system.
\begin{align}\label{formula:h_emb_el_ph_2}
\hat{H}^{\rm emb}_{\rm el-ph} = P^{\dagger}_{\rm ph}P^{\dagger}_{\rm el}\hat{H}_{\rm el-ph}P_{\rm el}P_{\rm ph}
\end{align}
\begin{figure}[t]
	\includegraphics[width=.5\textwidth]{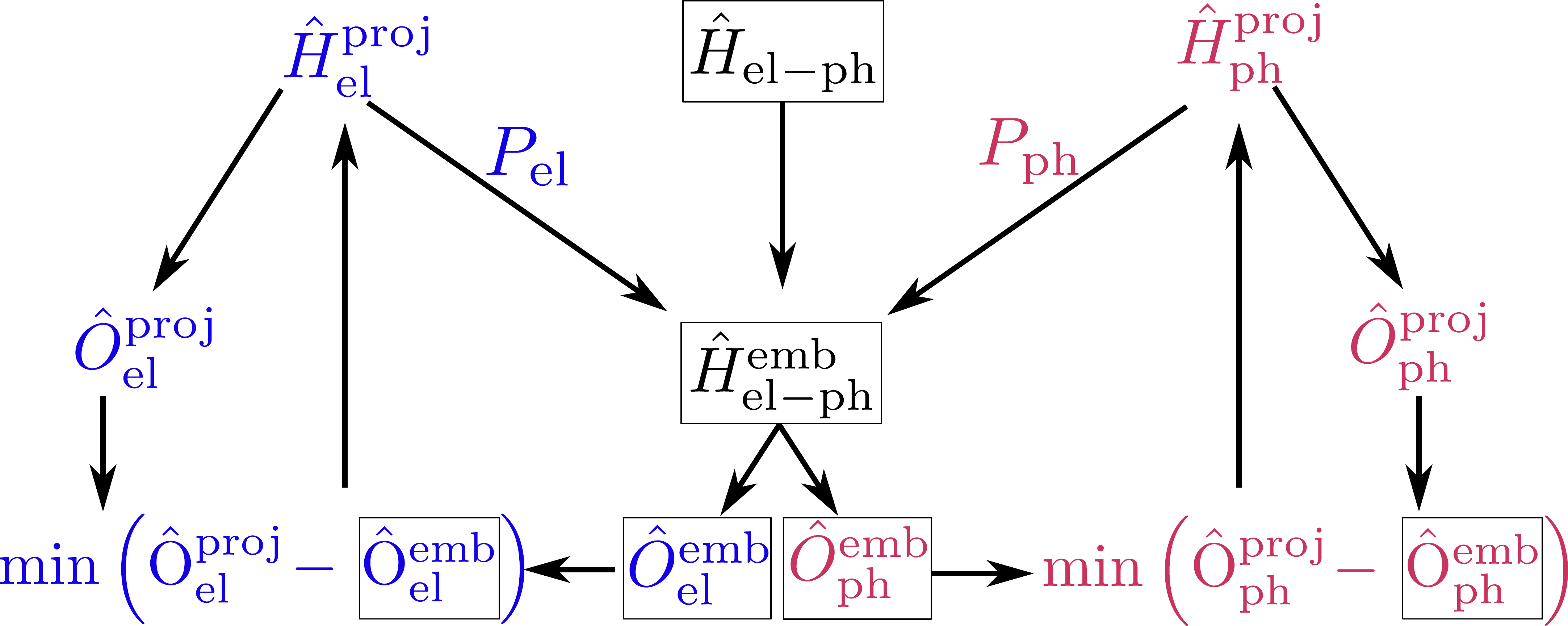}
    \caption{Visualization of the DMET procedure: from the purely electronic and the purely fermionic projected Hamiltonians $\hat{H}_{\rm el}^{\rm proj}$ and $\hat{H}_{\rm ph}^{\rm proj}$, we obtain the projections $P_{\rm el}$ and $P_{\rm ph}$. Applied to the full Hubbard-Holstein Hamiltonian $\hat{H}_{\rm el-ph}$ these yield the embedding Hamiltonian $\hat{H}^{\rm emb}_{\rm el-ph}$. In order to improve the projections $P_{\rm el}$ and $P_{\rm ph}$, we aim at making the electronic and phononic one-body properties of the interacting ($\hat{O}^{\rm emb}_{\rm ph}$, $\hat{O}^{\rm emb}_{\rm ph}$) and the non-interacting ($\hat{O}^{\rm proj}_{\rm ph}$, $\hat{O}^{\rm proj}_{\rm ph}$) systems as similar as possible. This is done by adding non-local potentials to the projected Hamiltonians that minimize the difference between the one-body observables of the interacting system and the non-interacting systems. When the new potentials are found, new projections can be calculated which yield a new embedding Hamiltonian. This procedure is repeated until the non-local potentials found do not differ up to an accuracy of $10^{-5}$ from the non-local potentials of the iteration before.}
    \label{pic:procedure}
\end{figure}
\\
The initial guess for the projection is not necessarily very good, as additionally to assuming a non-interacting active space for both the electrons as well as the phonons, it also assumes a product state between electronic and phononic degrees of freedom. \\ We self-consistently optimize the electronic and the phononic projection, where for the electronic case we proceed in the same manner as the purely electronic problem, as stated in Eq.~(\ref{formula:min_el}). For the phonons, we have to compare two properties as the initial guess for the projection Hamiltonian defined in Eq.~(\ref{formula:h_ph}) also has two terms
\begin{align}\label{formula:h_proj_ph}
\hat{H}^{\rm proj}_{\rm ph} = \hat{T}_{\rm ph} + \hat{V}_{\rm ph} + \hat{C}_{\rm ph} + \hat{W}_{\rm ph}.
\end{align}
While $\hat{V}_{\rm ph}$ depends on the phononic reduced one-particle density matrix $\langle\hat{a}_i^{\dagger}\hat{a}_j \rangle$, $\hat{W}_{\rm ph}$ depends on the shift of the phonons from zero $\langle\hat{a}^{\dagger}_i+\hat{a}_i\rangle$. The potentials are again found by minimizing the difference between the properties of the interacting and the noninteracting system:
\begin{align}\label{formula:min_ph}
\rm min \bigg| &\sum_{i,j\in \mathrm{imp}} \left< \Psi_{\rm emb}|a_i^{\dagger}a_j|\Psi_{\rm emb} \right> - \left<Z|a_i^{\dagger}a_j|Z\right> \nonumber\\
+ &\sum_{i\in \mathrm{imp}} \left< \Psi_{\rm emb}|\hat{a}^{\dagger}_i+\hat{a}_i|\Psi_{\rm emb} \right> - \left<Z|\hat{a}^{\dagger}_i+\hat{a}_i|Z\right> \bigg|,
\end{align}
where $\ket{Z}$ is the ground-state wave function of the Hamiltonian defined in Eq.~(\ref{formula:h_proj_ph}) and $\ket{\Psi_{\rm emb}}$ is the ground state wave function of the embedding Hamiltonian defined in Eq.~(\ref{formula:h_emb_el_ph_2}).\\
The whole DMET procedure is visualized in Figure \ref{pic:procedure}.\\
The embedding problem is then solved using MPS-DMRG\cite{white_density_1992,schollwock_density-matrix_2011,hubig_strictly_2015}. We obtain the optimal site ordering \cite{rissler_measuring_2006} from an initial approximate
calculation \cite{legeza}.
This site ordering is then used in a second higher-precision
calculation. In both cases, we construct the Hamiltonian as described in Ref.~\cite{hubig_generic_2017}. Electronic and phononic sites were kept separate and the DMRG3S algorithm\cite{hubig_strictly_2015} was used to achieve linear scaling with the relatively large local dimension.

\section{Phases of the one-dimensional Hubbard-Holstein Model}\label{section:obs}
\subsection{Physics}
In the one-dimensional Hubbard-Holstein model, three competing forces can be found: first, the electron hopping strength $t$, that leads to mobilization of the electrons and will put the system in a metallic phase. Second, the electron-electron interaction $U$ that, if dominant, leads to an immobilized Spin wave for the electronic degrees of freedom, that is, a Mott phase. Third, the electron-phonon coupling $g$, that, if dominant, leads to a Peierls phase, that is the position of the electrons on the lattice is distorted from the initial position, forming a charge density wave.\\
Due to strong quantum fluctuations of the phonons, the Peierls phase can be destroyed and can lead to a metallic phase, if the electron-electron interactions are not too strong to prevent this.
This is why we expect a distinct metallic phase when considering high phonon frequencies $\omega_0$ in comparison to the itineracy of the electrons $t$. In contrast, if the phonon  frequency is small compared to the electronic hopping, the metallic phase should, if existent, be smaller than in the anti-adiabatic limit.
\subsection{Observables monitoring the phase transition}
In order to describe the phase transition of the one dimensional Hubbard-Holstein model, we need to define observables that unambiguously show which phase the system is in.\\
We choose here three different observables, namely the double occupancy $\left<n_{i\uparrow}n_{i\downarrow}\right>$, the electronic density difference between neighboring sites $\left<n_i\right>-\left<n_{i+1}\right>$ and the energy gap $\Delta c$ defined in Eq.~\ref{formula:gap}.\\
The double occupancy and the electronic density difference between neighboring sites are local properties and can simply be calculated on one arbitrary (impurity) site. Unfortunately though, the double occupancy only gives a rough estimate of the phase and the electronic density difference between neighboring sites only indicates the transition to the Peierls phase as in the Mott phase, the electronic density stays homogeneous.\\
The energy gap $\Delta c$ indicates unambiguously whether the system is in the metallic phase (where the gap is zero) or in an insulating state, which can be either Mott or Peierls (where the gap is finite). Unfortunately though, it cannot be simply defined locally but is a global property of the whole system for different particle numbers
\begin{align}\label{formula:gap}
\Delta c = 2\cdot E^{\rm N/2}_0 - E^{\rm N/2-1}_0 - E^{\rm N/2+1}_0,
\end{align}
where $E^{N/2}_0$ is the ground state energy of the Hamiltonian for half filling, $E^{N/2-1}_0$
 is the ground state energy of the system for half filling minus one and $E^{N/2+1}_0$ is the energy of the system for half filling plus one.\\
 As our DMET calculation has only been implemented for closed shell systems, we have to approximate the calculation of the gap. Instead of doing three calculations with different particle numbers, we consider our "sophisticated mean field" Hamiltonian
 \begin{align}
 \hat{H}^{\rm proj}_{\rm el-ph} = \hat{T}_{\rm el} + \hat{V}_{\rm el} + \hat{T}_{\rm ph} + \hat{V}_{\rm ph} + \hat{C}_{\rm ph} + \hat{W}_{\rm ph}
 \end{align}
 which is optimized to have similar one-body properties as the interacting Hamiltonian. We calculate the (one-body) spectrum of this Hamiltonian by diagonalizing it and then approximate the gap by defining
 \begin{align}
  \Delta c &= 2\cdot E^{\rm N/2}_0 - E^{\rm N/2-1}_0 - E^{\rm N/2+1}_0 \nonumber\\
 &= \epsilon_{\rm N/2}-\epsilon_{\rm N/2+1},
 \end{align}
 where
 \begin{align}
 \hat{H}^{\rm proj}_{\rm el-ph}\ket{\Phi} &= \epsilon_i\ket{\Phi};\,\,\,\,
 E^{M}_0 = \sum_{i=1}^{M} \epsilon_i.
 \end{align}
 \subsection{Parameters}
 The phase transition depends on the itineracy of the electrons $(\propto t)$, the electron-electron repulsion $(\propto U)$, the electron-phonon interaction $(\propto g)$ and the relative velocity of the phononic degrees of freedom with respect to the electrons. This is why, we introduce the adiabaticity ratio
  \begin{align}
 \alpha = \frac{\omega_0}{t}
 \end{align}
 accounting for the relation between the kinetic hopping energy of the electrons $t$ and the frequency of the phonons $\omega_0$. We also decide to discuss our results in terms of dimensionless coupling constants:
 \begin{align}
 u=\frac{U}{4t},\,\,\,\,\lambda=\frac{g^2}{2t\omega_0}
 \end{align}
 
\section{Results}\label{section:results}
In the following section, we discuss the results of our DMET calculation when solving the Hubbard-Holstein model Hamiltonian. First, in section \ref{section:dmet}, we describe the quantum phase transitions in the anti-adiabatic $(\alpha = 5.0)$ and in the adiabatic $(\alpha=0.5)$ limit. Afterwards, in section \ref{section:dmrg}, we compare the DMET results with results obtained from a pure real-space DMRG calculation on the original chain, using the code presented in \cite{hubig_strictly_2015}. Finally, in section \ref{section:bo}, we investigate the importance of the quantum nature of the phonons by comparing to DMET results with the Born Oppenheimer (BO) approximation, i.e. by regarding the phonons as the vibrations of classical ions.\\
Both for the DMET as well as for the DMRG calculation, we did a finite size scaling whose discussion can be found in the appendix (section \ref{section:appendix}). For the DMET calculations, this includes a finite size scaling with respect to the total system size as well as a finite size scaling with respect to the impurity size. In the DMRG calculation, only the finite size scaling of the total system size needs to be considered.\\
Throughout this paper, in the DMET calculation we consider 8 phononic modes per lattice site and a bond dimension of 2000 for the DMRG impurity solver. In the DMRG reference calculations, we consider a maximal bond dimension of 4000 and up to 10 phononic modes per lattice site.
\subsection{DMET results}\label{section:dmet}
\subsubsection{Anti-adiabatic limit}\label{section:anti-adi}
\begin{figure}[t]
	\includegraphics[width=.5\textwidth]{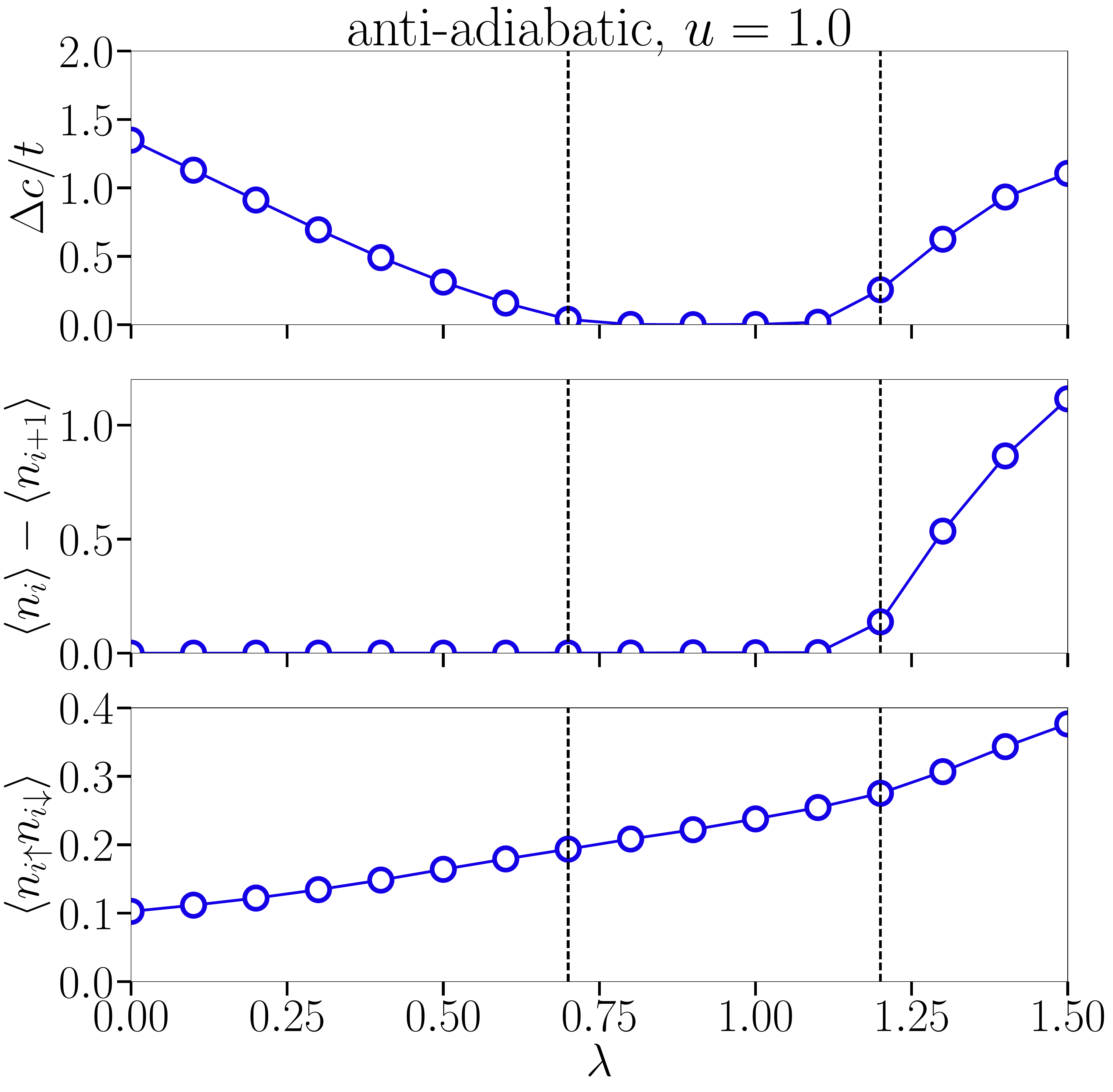}
    \caption{Energy gap $\Delta c/t$, density difference of the electrons between neighboring sites $\left<n_i\right>-\left<n_{i+1}\right>$ and double occupancy $\left<n_{i\uparrow}n_{i\downarrow}\right>$ for the anti-adiabatic limit $\alpha = 5.0$ and an electron-electron coupling of $u=1.0$ for different electron-phonon couplings $\lambda$. For $0 \le \lambda \le 0.7$, a Mott phase is observed, which changes into a metallic phase for $0.7 \le \lambda \le 1.1$. Above coupling values of $1.1 \le \lambda$, we observe a Peierls phase.} 
    \label{graph:anti_dmet}
\end{figure}

In Figure \ref{graph:anti_dmet}, we plot the energy gap $\Delta c/t$, the electronic density difference between neighboring sites $\left<n_i\right>-\left<n_{i+1}\right>$ and the double occupancy $\left<n_{i\uparrow}n_{i\downarrow}\right>$ (as defined in section \ref{section:obs}) in the anti-adiabatic limit $(\alpha =5.0)$ for an electron-electron repulsion of $u=1.0$ and for different electron-phonon coupling strengths $\lambda$.

For all three observables, we observe a Mott phase for $0 \le \lambda \le 0.7$. With growing $\lambda$, we indeed observe a distinct metallic phase $(0.7 \le \lambda \le 1.1)$ which is followed by a Peierls phase for $1.1 \le \lambda$.

\subsubsection{Adiabatic limit}\label{section:adi}
\begin{figure*}[t]
	\includegraphics[width=\textwidth]{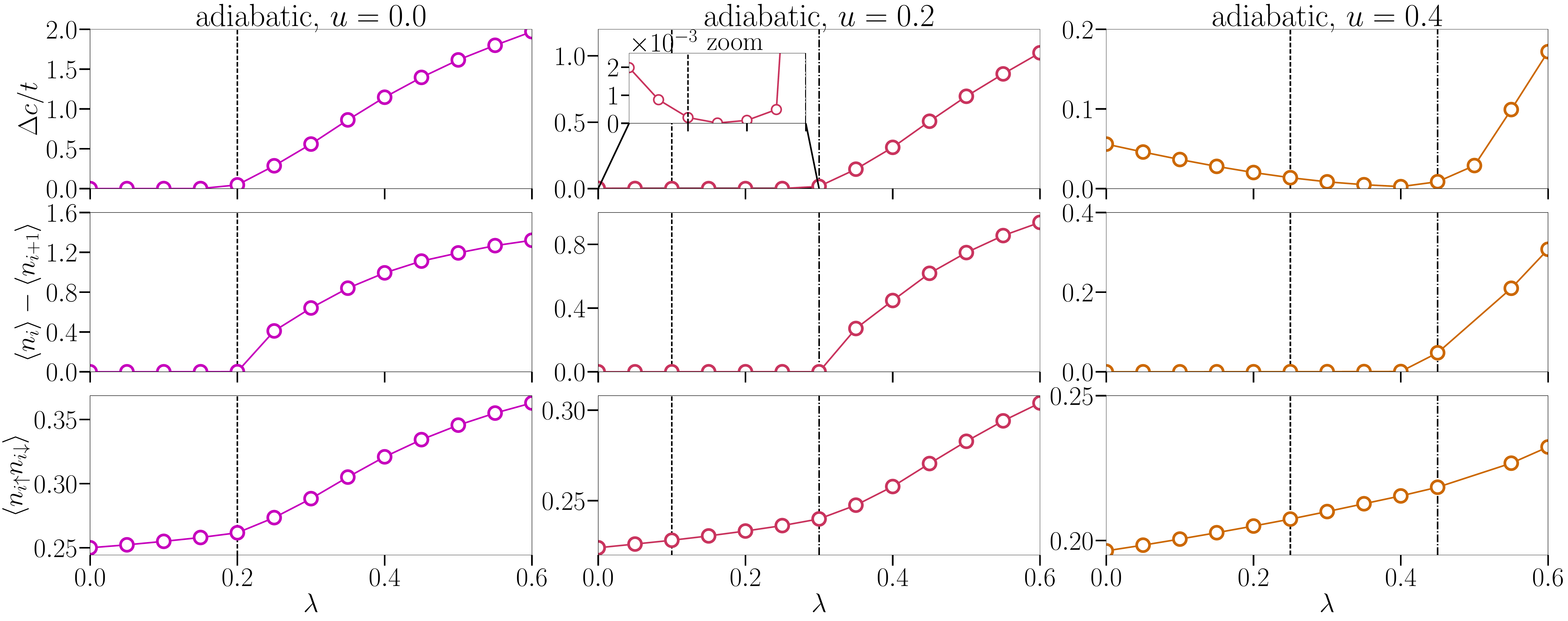}
    \caption{Energy gap $\Delta c/t$, density difference of the electrons between neighboring sites $\left<n_i\right>-\left<n_{i+1}\right>$ and double occupancy $\left<n_{i\uparrow}n_{i\downarrow}\right>$ for the adiabatic limit $\alpha = 0.5$ and three different electron-electron couplings, $u=0.0$, $u=0.2$ and $u=0.4$ for different electron-phonon couplings $\lambda$. For $u=0.0$ (absent electron-electron coupling), we do not observe any Mott phase but a direct transition from the metallic to the Peierls phase at $\lambda = 0.2$. For a value of $u=0.2$, the Mott phase exists for values of $0\le \lambda \le 0.1$ but the gap is very small. For values $0.1\le \lambda \le 0.3$, a metallic phase can be observed, followed by a Peierls phase for $\lambda \ge 0.3$. For bigger electron-electron couplings $(u=0.4)$ the energy gap indicating the Mott phase (from $0\le\lambda \le 0.25$) gets more pronounced. Within the error bars, we get a metallic phase for the range of $(0.25\le \lambda \le 0.45)$ which stays the same for this $u$-value while the gap due to the Peierls phase for $0.45 \le \lambda $ becomes less pronounced (but still visible).}
    \label{graph:adi_dmet}
\end{figure*}

The occurence of a pronounced metallic phase in the anti-adiabatic limit was to be expected; it is however not clear whether this phase also occurs for all electron-electron interaction strength $u$ in the adiabatic limit, where the phonon frequency is small in comparison to the electronic hopping and thus, the quantum fluctuations of the phonons are suspected to be smaller.\\
In Figure \ref{graph:adi_dmet}, we again show the energy gap $\Delta c/t$, the electronic density difference between neighboring sites $\left<n_i\right>-\left<n_{i+1}\right>$ and the double occupancy $\left<n_{i\uparrow}n_{i\downarrow}\right>$ (as defined in section \ref{section:obs}) in the adiabatic limit $(\alpha =0.5)$ for different electron-electron repulsions, $(u=0.0;0.2;0.4)$ and different electron-phonon coupling strengths $\lambda$.

\begin{figure}[t]
	\includegraphics[width=.5\textwidth]{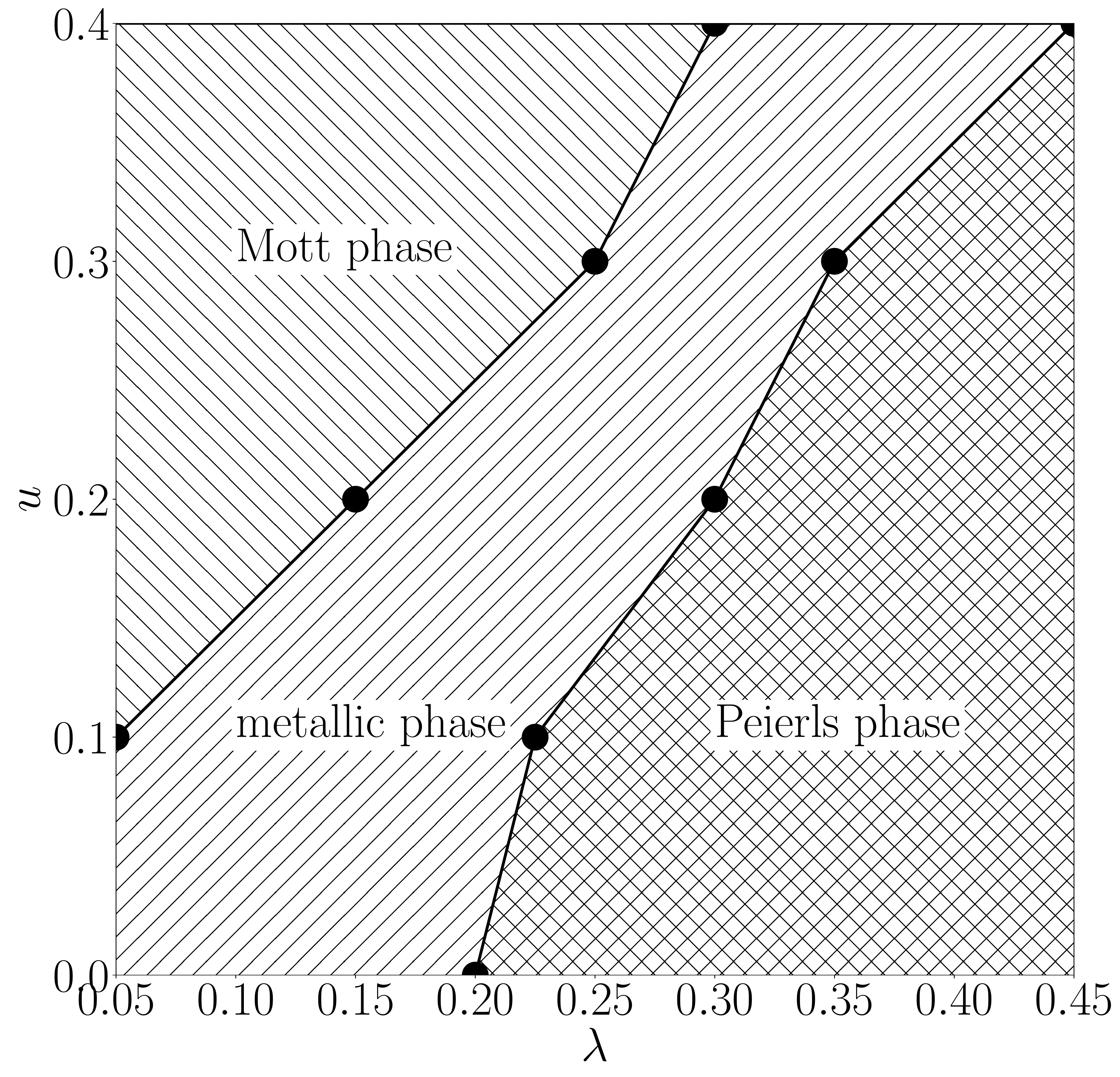}
    \caption{Phase diagram for the adiabatic limit $(\alpha=0.5)$ of the Hubbard-Holstein model. For different electron-electron coupling values $u$ and different electron-phonon coupling values $\lambda$, the phase of the model at these parameters is indicated. While not existent at all for $u=0$, the Mott phase gets more and more pronounced with growing $u$ and small $\lambda$ values. The Peierls phase, while always existing in this range, needs higher electron-phonon coupling strength to occur when the electron-electron interactions are also growing bigger.}
    \label{graph:phase_diag_adi}
\end{figure}

When the electron-electron interaction is absent, we do not observe a Mott phase, but a direct transition from the metallic to the Peierls phase at $\lambda=0.2$. This result is as expected as the Mott phase is driven by the electron-electron interaction and therefore cannot occur in this limit.\\
For a small electron-electron interaction, $u=0.2$, the Mott phase exists for very small electron-phonon interactions $0 \le \lambda \le 0.1$. The gap indicating the Mott phase though is very small in comparison to the gap that indicates the pronounced Peierls phase for $0.3 \le \lambda$. Between the Mott and the Peierls, we observe a small metallic phase for electron phonon coupling values of $0.1\le \lambda \le 0.3$.\\
When considering bigger electron-electron interactions $u=0.4$, the size of the gap indicating the Mott gap grows considerably, as does the range of the Mott phase: for $0 \le \lambda \le 0.25$, we observe a Mott phase, followed again by a narrow metallic phase for $0.25 \le \lambda \le 0.45$. Afterwards, we observe a Peierls phase, whose gap is less pronounced than for lower $u$, but still clearly visible.\\
Our results for the adiabatic limit of the Hubbard-Holstein model are summarized in the phase diagram shown in Figure \ref{graph:phase_diag_adi}. 

We observe that the Mott phase, while not existent at all for $u=0.0$, grows more and more pronounced for growing electron-electron interaction values $u$. The range of the metallic phase stays approximately constant for different $u$-values, but shifts from small values for electron-phonon interaction $\lambda$ to intermediate values. The Peierls phase is getting less pronounced for growing $u$.

\subsection{Comparison DMET and DMRG calculations}\label{section:dmrg}
\begin{figure}[t]
	\includegraphics[width=.5\textwidth]{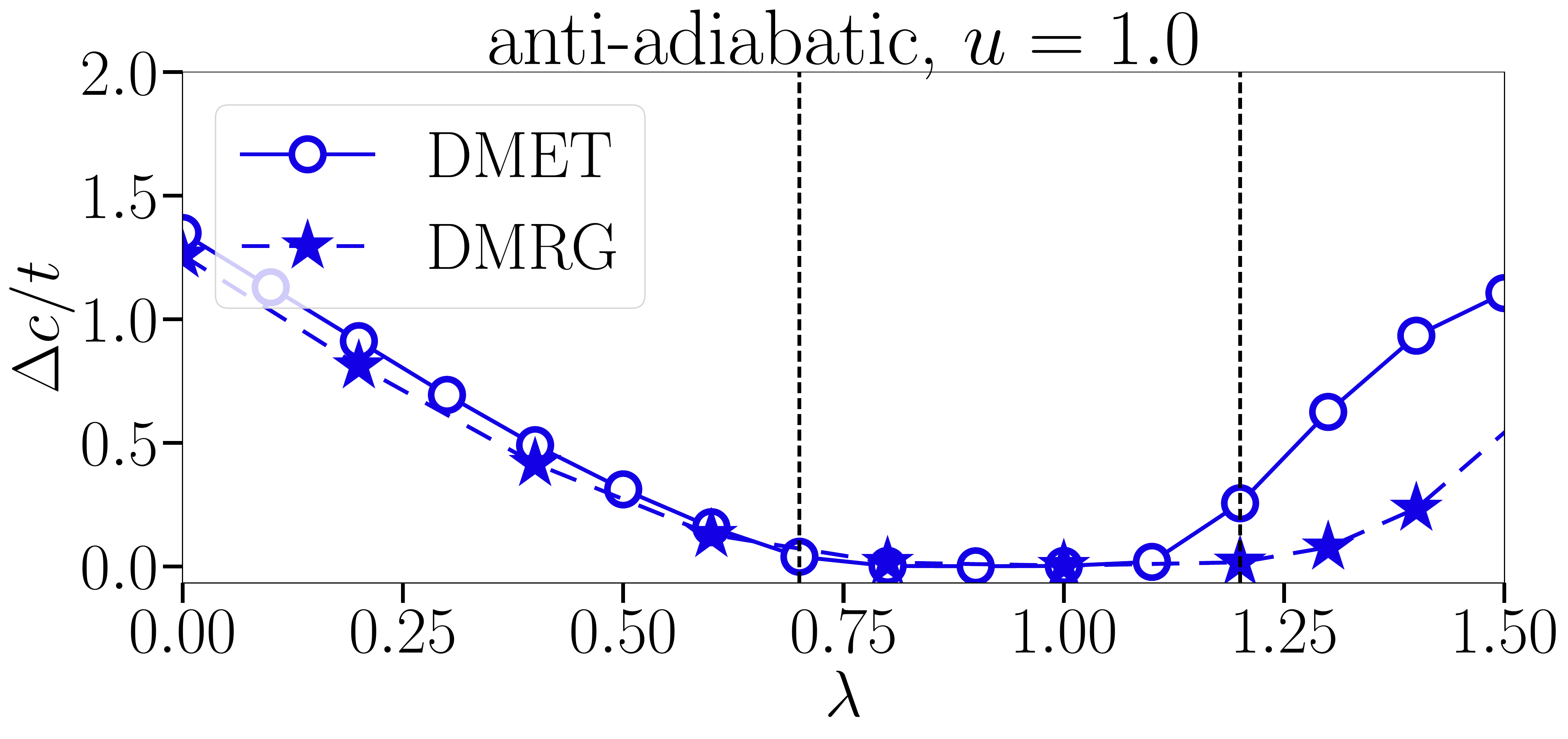}
    \caption{Comparison of the energy gap $\Delta c/t$ for the DMRG and the DMET calculation in the anti-adiabatic limit $\alpha=5.0$. Plotted are different electron-phonon coupling values $\lambda$ for a constant electron-electron coupling of $u=1.0$. We observe a quantitative agreement in the position of the phase transitions between Mott and metallic phase at $\lambda=0.7$, the phase transition between metallic and Peierls phase is at $\lambda=1.2$ for the DMET calculation while it is at $\lambda=1.3$ for the DMRG calculation. While the size of the Mott gap agrees quantitatively, the size of the Peierls gap is only in qualitative agreement.}
    \label{graph:anti_dmrg_dmet}
\end{figure}

\begin{figure}[b]
	\includegraphics[width=.5\textwidth]{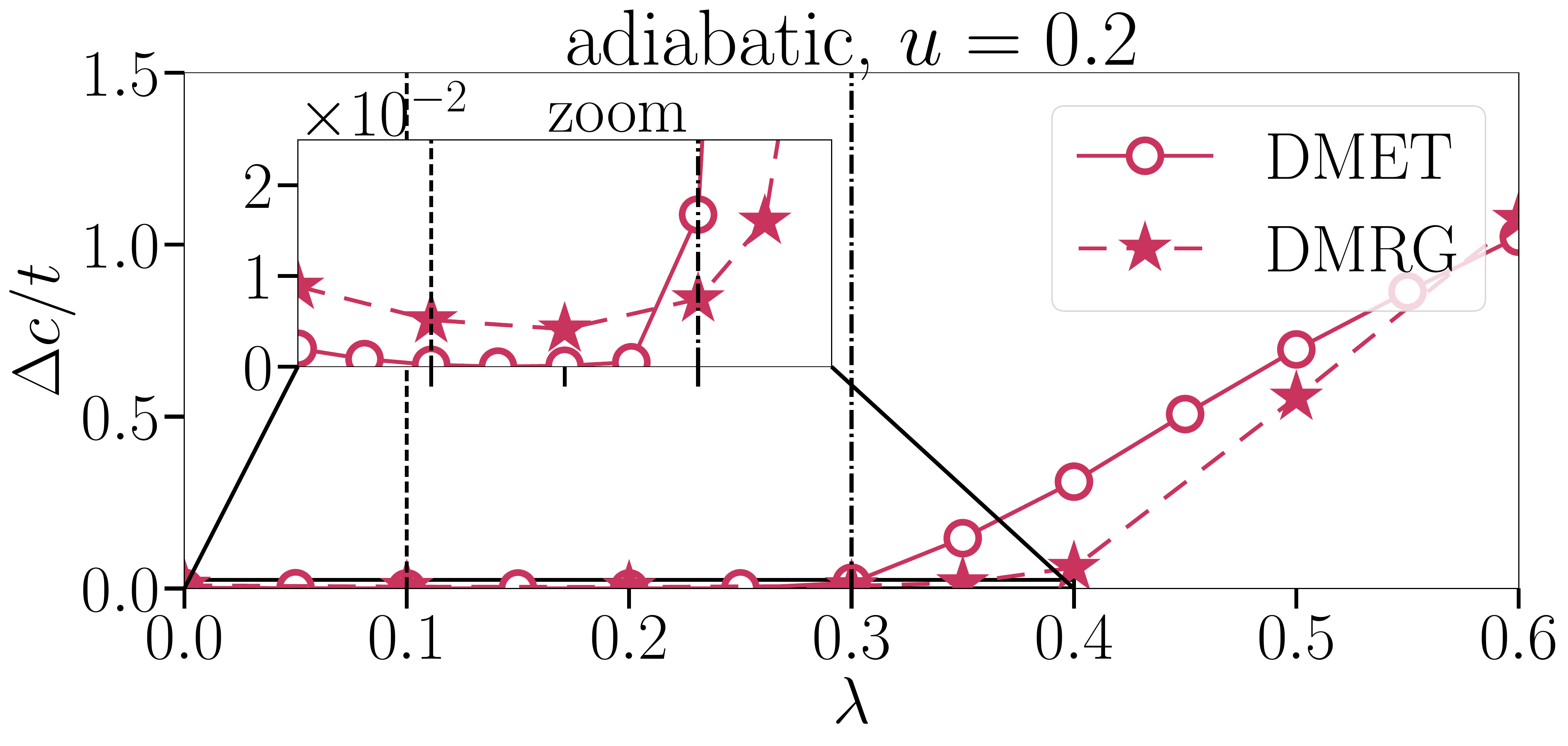}
    \caption{Comparison of the energy gap $\Delta c/t$ for the DMRG and the DMET calculation in the adiabatic limit $\alpha=0.5$. Plotted are different electron-phonon coupling values $\lambda$ for a constant electron-electron coupling of $u=0.2$. We observe a quantitative agreement in the position of the phase transitions between Mott and metallic phase at $\lambda=0.1$ and between metallic and Peierls phase at $\lambda=0.3$. For the Mott phase and the Peierls phase both, the size of the gap is only in qualitative agreement.}
    \label{graph:adi_dmrg_dmet}
\end{figure}

We benchmark the accuracy of DMET against results obtained using the DMRG method.
The results are obtained with the \textsc{SyTen} library, that for this purpose was expanded to be able to treat coupled fermion-boson systems. 
We compare the DMRG and the DMET results for both the anti-adiabatic limit ($\alpha = 5.0$) and the adiabatic ($\alpha = 0.5$) limit.\\
In Figure \ref{graph:anti_dmrg_dmet}, we compare the DMRG and the DMET results for the anti-adiabatic limit ($\alpha = 5.0$) and an electron-electron repulsion of $u=1.0$. 
Up to an electron-phonon coupling value of $\lambda=1.2$, we observe a quantitative agreement of the energy gap, although different approximations where made to calculate this property (while in the DMET calculation, we only take the HOMO-LUMO gap of the mean field system, in the DMRG calculation we calculate the energy gap for systems with half, half plus one and half minus one filling, as explained in section \ref{section:obs}). While for higher values of $\lambda$, the actual value of the gap differs, the point of the quantum phase transition is predicted equivalently for the DMRG and the DMET calculation. For a value of $\lambda=1.3$, the gap measured by the DMET calculation abruptly increases, while it only increases slightly for the DMRG calculation. Reasons for this discrepancy can be either due to the DMET or the DMRG treatment of the system: In the DMET treatment, we consider the gap only for the mean field calculation which overestimates the Peierls phase. In the DMRG treatment, it is not feasible to treat more than 10 phononic basis function per site which could lead to an under-estimation of the Peierls phase in the DMRG calculation.\\
In Figure \ref{graph:adi_dmrg_dmet}, we compare the DMRG and the DMET results for the adiabatic limit ($\alpha = 0.5$) and an electron-electron repulsion of $u=0.2$. 
While the position of the phase transitions both between the Mott and the metallic phase as well as between the metallic and the Peierls phase agree quantitatively, the actual sizes of the gaps only agree qualitatively: In the Mott and metallic phase, the gap measured by the DMRG calculation is bigger than the gap measured in the DMET calculation. While in the metallic phase, the DMET gap closes up to a value of $10^{-4}$, the value stays at a value of $5\cdot 10^{-3}$ in the DMRG calculation. For the metallic phase, these deviations are within the error limit of the finite size scaling presented in section \ref{section:finite_size} although, as stated before, different approximations where made in order to obtain the energy gap.\\
In the Peierls phase, as also observed in the anti-adiabatic limit, the size of the gap in the DMET calculation is bigger as in the DMRG calculation. As already discussed before, this can have its origin either in the mean field nature of the calculation of the DMET gap or in the cutting of the Fock space in the DMRG calculation.

\subsection{The Born Oppenheimer approximation}\label{section:bo}
\begin{figure}[t]
	\includegraphics[width=.5\textwidth]{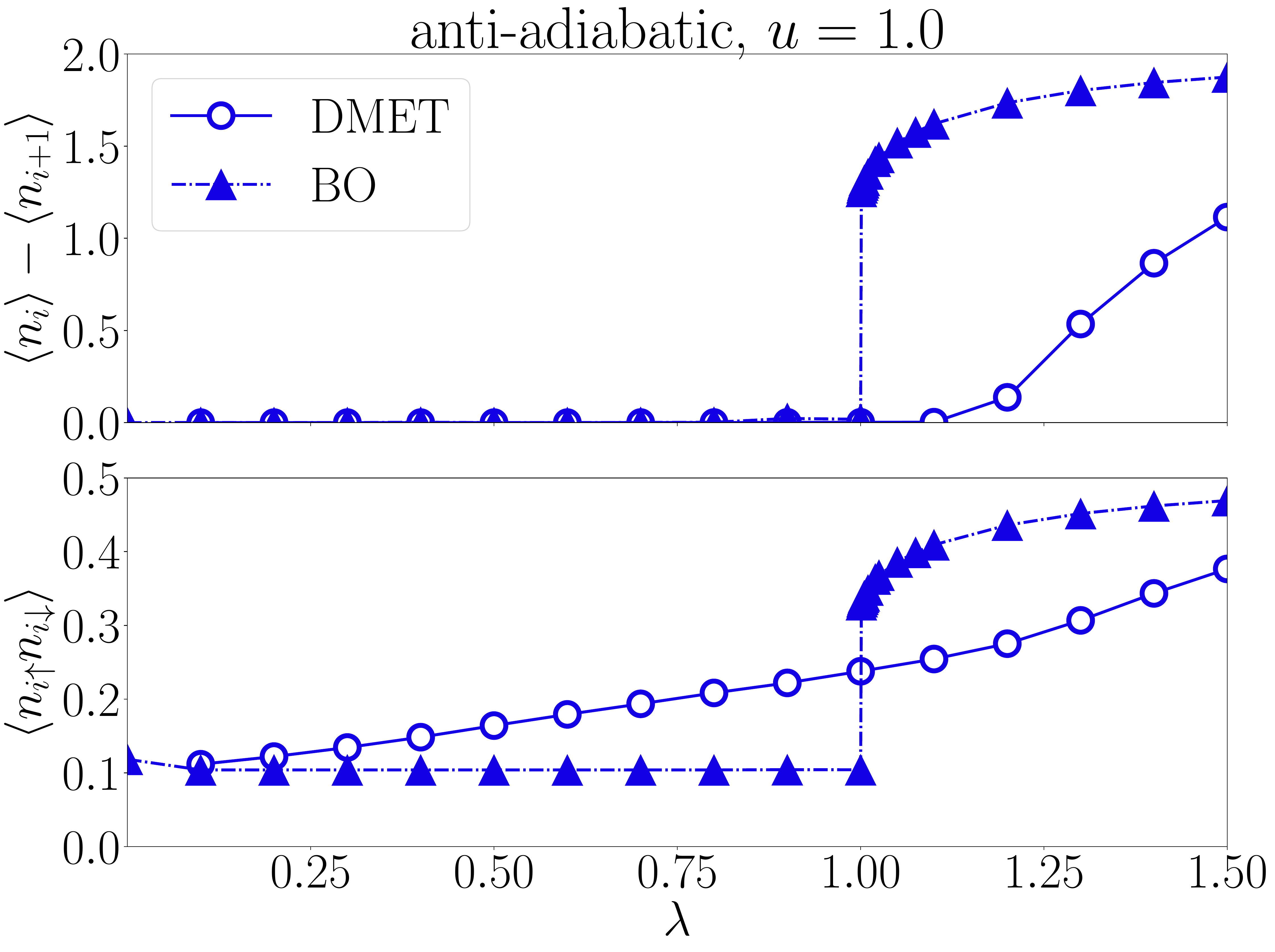}
        \caption{Comparison of the density difference of the electrons between neighboring sites $\left<n_i\right>-\left<n_{i+1}\right>$ and the double occupancy $\left<n_{i\uparrow}n_{i\downarrow}\right>$ in the anti-adiabatic limit for the fully quantum mechanical treatment and the BO approximation of the Hubbard-Holstein model. Both calculations are performed with DMET. While, for the fully quantum mechanical treatment the position of the phase transition is grasped quantitatively, both the nature as well the position of the gap is predicted falsely in the BO approximation, predicting a first-order phase transition as well as an earlier occurence of the transition. Parameters are $\alpha = 5.0$, $u=1.0$, $t_{\rm elec}=t_{\rm phon}=1$, $N_{\rm imp}=6$, $N_{\rm phon} = 8$.}
        \label{graph:anti_bo_dmet}
\end{figure}

The Hamiltonian has so far been written in second quantized form, but it can be equivalently also written as:
\begin{align}
	\hat{H}_{\rm{el-ph}}=t\sum_{<i,j>,\sigma}\hat{c}_{i\sigma}^{\dagger}\hat{c}_{j\sigma} + U \sum_i \hat{n}_{i\uparrow}\hat{n}_{i\downarrow} \nonumber\\
    + \omega_0\sum_i\left(\frac{\omega_0}{2}\hat{x}_i^2+\frac{1}{2\omega_0}\hat{p}_i^2\right) + gn_i\sqrt{2\omega_0}\hat{x}_i.
\end{align}
In contrast to Eq.~(\ref{formula:hub_hol}), here the bosonic degrees of freedom are not considered in terms of phonons, but, in terms of the distortion from the initial position $\hat{x}_i$ of the ions. $\hat{p}_i$ is the momentum of the ions.\\
In the BO approximation, we assume the ions to be classical particles as, due to their higher mass, they are moving much slower than the electrons. Thus we can neglect their kinetic energy which yields:
\begin{align}
\hat{H}_{\rm{el-ph}}=t\sum_{<i,j>,\sigma}\hat{c}_{i\sigma}^{\dagger}\hat{c}_{j\sigma} + U \sum_i \hat{n}_{i\uparrow}\hat{n}_{i\downarrow}\nonumber\\
+\sum_i \frac{\omega_0^2}{2}\hat{x}_i^2 + \xcancel{\sum_i \frac{1}{2}\hat{p}_i^2} + \sum_i gn_i\sqrt{2\omega_0}\hat{x}_i.
\end{align}
Here, the remaining ionic term, $\sum_i \frac{\omega_0^2}{2}\hat{x}_i^2$, purely depends on the distortion of the ions and can be treated as an external parameter. We treat the BO Hamiltonian with purely electronic DMET and optimize the distortion of the ions $\hat{x}_i$ to minimize the total energy. 
In Figure \ref{graph:anti_bo_dmet}, we compare the double occupancy $\left<n_{i\uparrow}n_{i\downarrow}\right>$ and the distortion of the electronic density $\left<n_i\right>-\left<n_{i+1}\right>$ for the BO system and the fully quantum mechanical system in the anti-adiabatic limit $(\alpha =5.0)$ and for an electron-electron repulsion of $u=1.0$. 

\begin{figure}[t]
	\includegraphics[width=.5\textwidth]{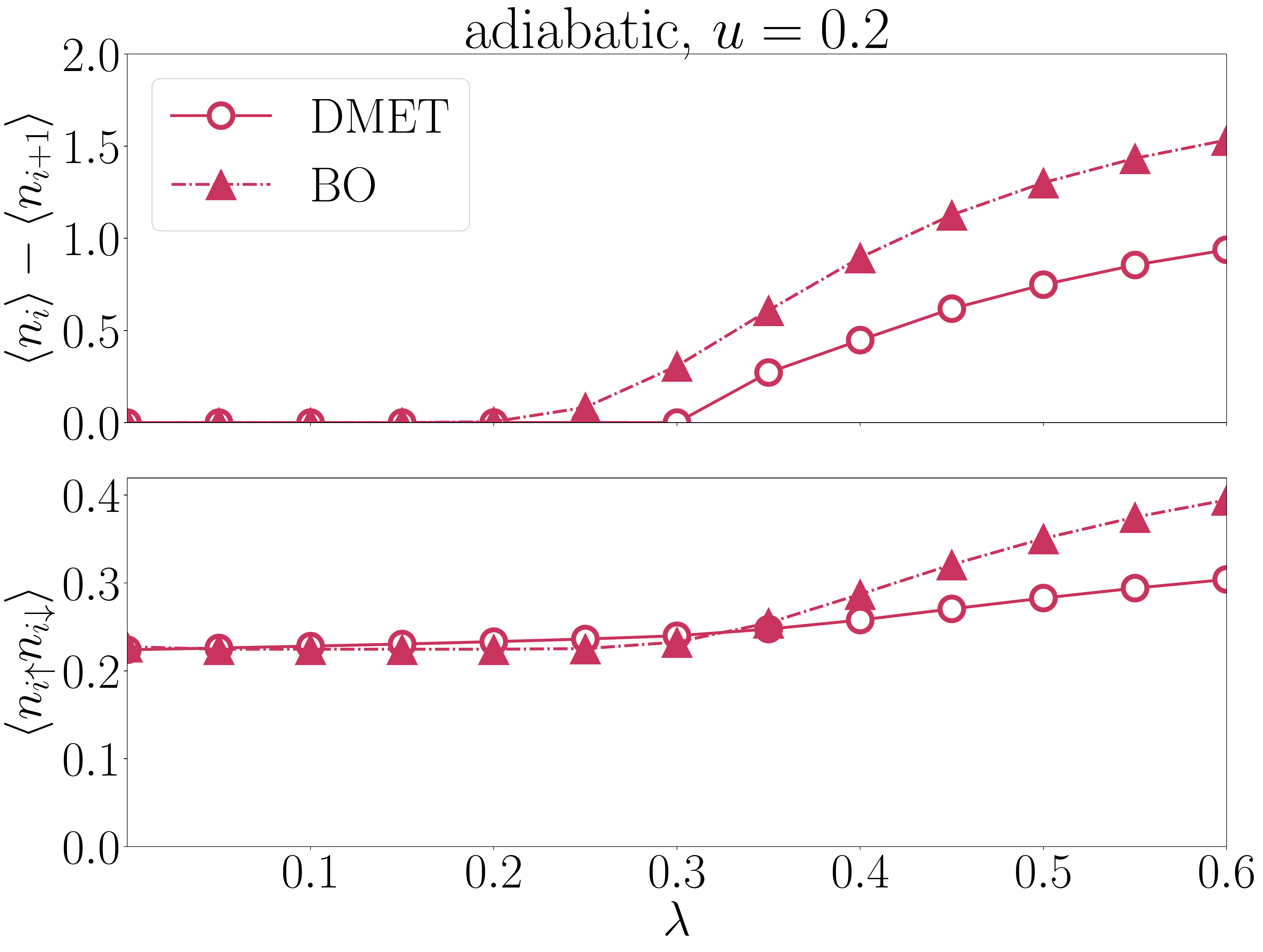}
    \caption{Comparison of the density difference of the electrons between neighboring sites $\left<n_i\right>-\left<n_{i+1}\right>$ and the double occupancy $\left<n_{i\uparrow}n_{i\downarrow}\right>$ in the adiabatic limit for the fully quantum mechanical treatment and the Born Oppenheimer approximation of the Hubbard-Holstein model. Both calculations are performed with DMET. While for the fully quantum mechanical treatment the position of the phase transition is grasped quantitatively, it is predicted too early with the Born Oppenheimer method. Parameters are $\alpha = 0.5$, $u=0.2$, $t_{\rm elec}=t_{\rm phon}=1$, $N_{\rm imp}=6$, $N_{\rm phon} = 8$.}
    \label{graph:adi_bo_dmet}
\end{figure}

We observe that for both observables, the Born-Oppenheimer description of the phase transition is not accurate. While in the full quantum mechanical model, the transition between metallic and Mott phase occurs for a value of $\lambda=1.1$, in the Born-Oppenheimer model, this transition already occurs for $\lambda = 1.0$. Additionally, the actual phase transition is of second order, while the Born-Oppenheimer treatment predicts a phase transition of first order.\\
In Figure \ref{graph:adi_bo_dmet}, we again compare the full quantum-mechanical treatment with the BO approximation, this time for the adiabatic limit and an electron-electron repulsion of $u=0.2$. While still not accurate (the phase transition is predicted too early, at $\lambda=0.25$ (BO) instead of $\lambda=0.3$ (full)), at least the qualitative nature of the phase transition as being of second order is grasped.

This result confirms our expectation that in order to treat the quantum phase transitions of the Hubbard-Holstein model, both the quantum mechanical nature of the electrons as well as of the phonons needs to be taken into account. Especially when the phononic frequency if higher than the electronic kinetic hopping, the BO approximation, which assumes the phonons to be moving much slower than the electrons, fails.

\section{Conclusions and outlook}\label{section:conclusion}
In conclusion, we have benchmarked the density-matrix embedding theory against density-matrix renormalization group results for the 
one-dimensional Hubbard-Holstein model.

We have demonstrated excellent agreement not only for groundstate energies but more notably of excitation gaps and phase diagrams between DMET and DMRG. 

An important prospect of DMET for the electron-boson problem lies in its possible extensions to electron-photon systems. Notably, recent efforts towards cavity quantum-electrodynamical engineering of materials properties \cite{laussy_exciton-polariton_2010,cotlet_superconductivity_2016,sentef_cavity_2018,schlawin_cavity-mediated_2018,mazza_superradiant_2018,curtis_cavity_2018,kiffner_manipulating_2018,allocca_cavity_2018} have been made.
We envision that these developments will open up a whole new field in which efficient methods able to deal with correlated electron-boson lattice systems from weak to strong coupling are urgently needed. Our benchmark study helps pave the way to these new endeavors. 

\section{Acknowledgements}
We would like to acknowledge helpful discussions with Garnet K.~Chan.
C.H. acknowledges funding through ERC Grant No.742102 QUENOCOBA.
M.A.~S. acknowledges financial support by the DFG through the Emmy Noether programme (SE 2558/2-1).
T. E.~R. is grateful for the kind hospitality of Princeton University, where a part of this project was carried out. 
U. M. acknowledges funding by the IMPRS-UFAST.
A.  R.  acknowledges financial support by the European Research Council  (ERC-2015-AdG-694097).  The Flatiron Institute is a division of the Simons Foundation.

\section{Appendix}\label{section:appendix}
\subsection{Finite size scaling}\label{section:finite_size}
\begin{figure}[t]
	\includegraphics[width=.5\textwidth]{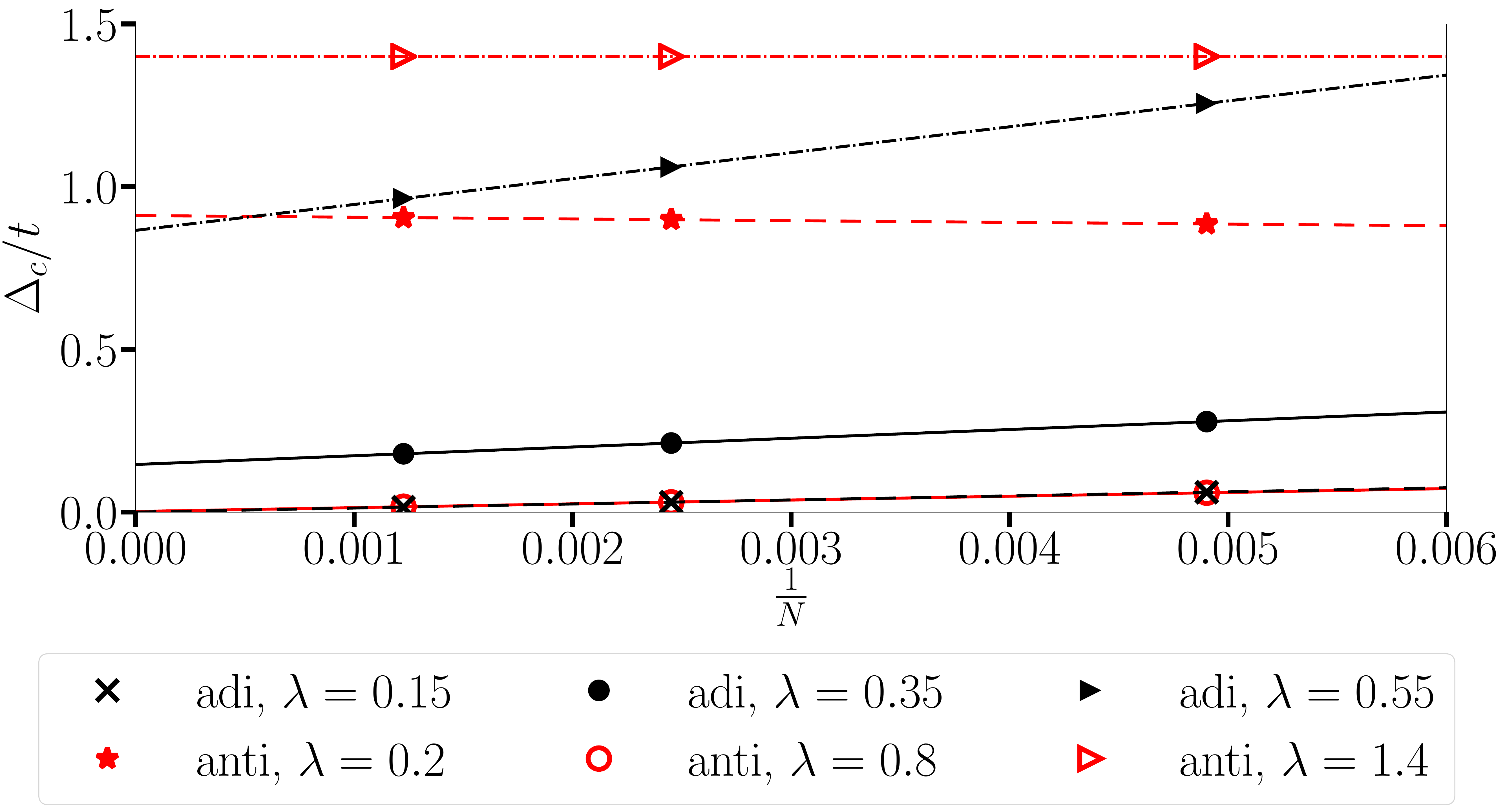}
    \caption{Finite size scaling for the energy gap in the DMET calculation. We show  some examples, both for the adiabatic limit $(\lambda=0.1;0.3;0.5$ and $u=0.2)$ as well as for the anti-adiabatic limit $(\lambda=0.2;0.8;1.4$ and $u=1.0)$. The extrapolation is done with system sizes of $N=204;408;816$. The scaling is linear, making it possible to remove finite size effects.}
    \label{graph:finite_size_dmet_gap}
\end{figure}

\begin{figure}[t]
	\includegraphics[width=.49\textwidth]{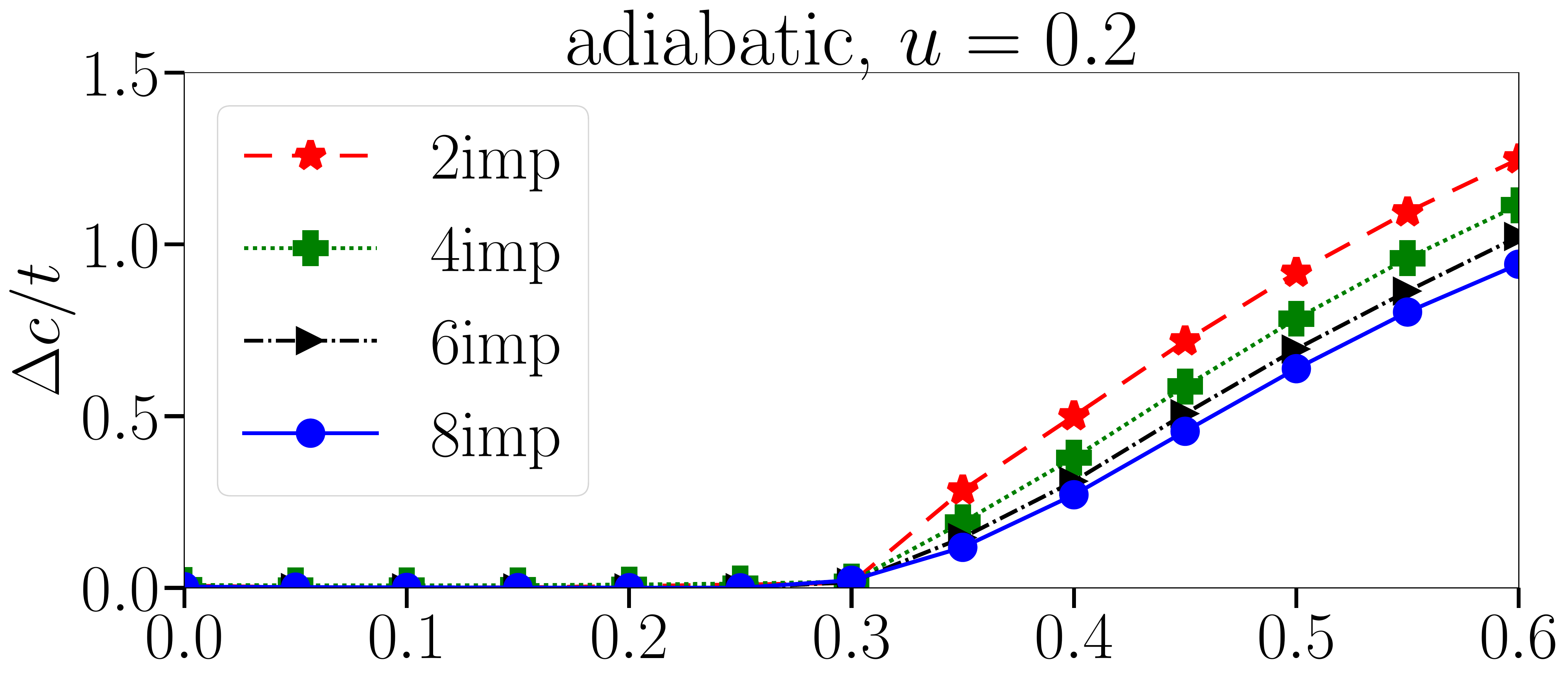}
	\includegraphics[width=.49\textwidth]{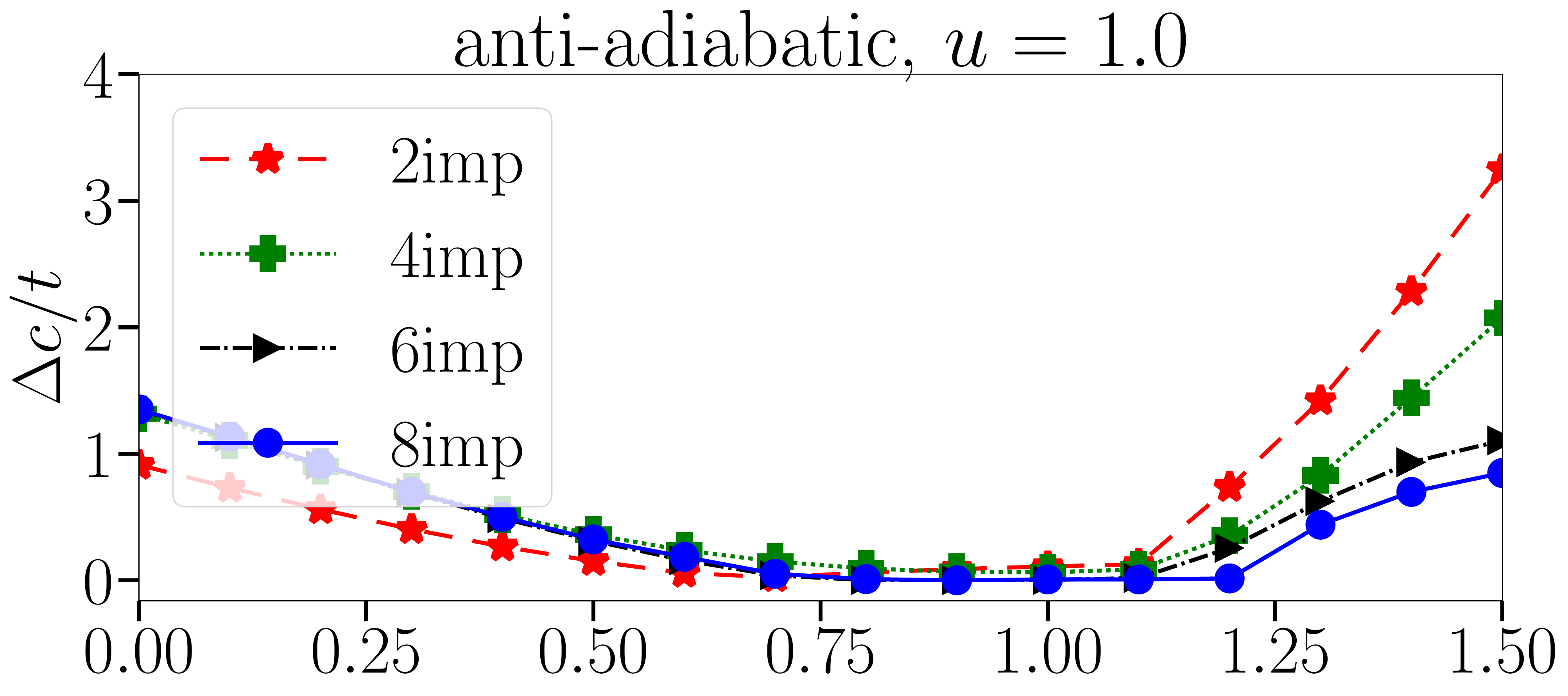}
	\caption{Scaling of the energy gap with growing impurity sizes for the adiabatic limit $(\lambda=0.1;0.3;0.5$ and $u=0.2)$ as well as for the anti-adiabatic limit $(\lambda=0.2;0.8;1.4$ and $u=1.0)$. We see that the discrepancy between the results gets smaller for growing impurity sizes but that the scaling is nonlinear.}
	\label{graph:finite_size_imp}
\end{figure}

\begin{figure}[t]
	\includegraphics[width=.5\textwidth]{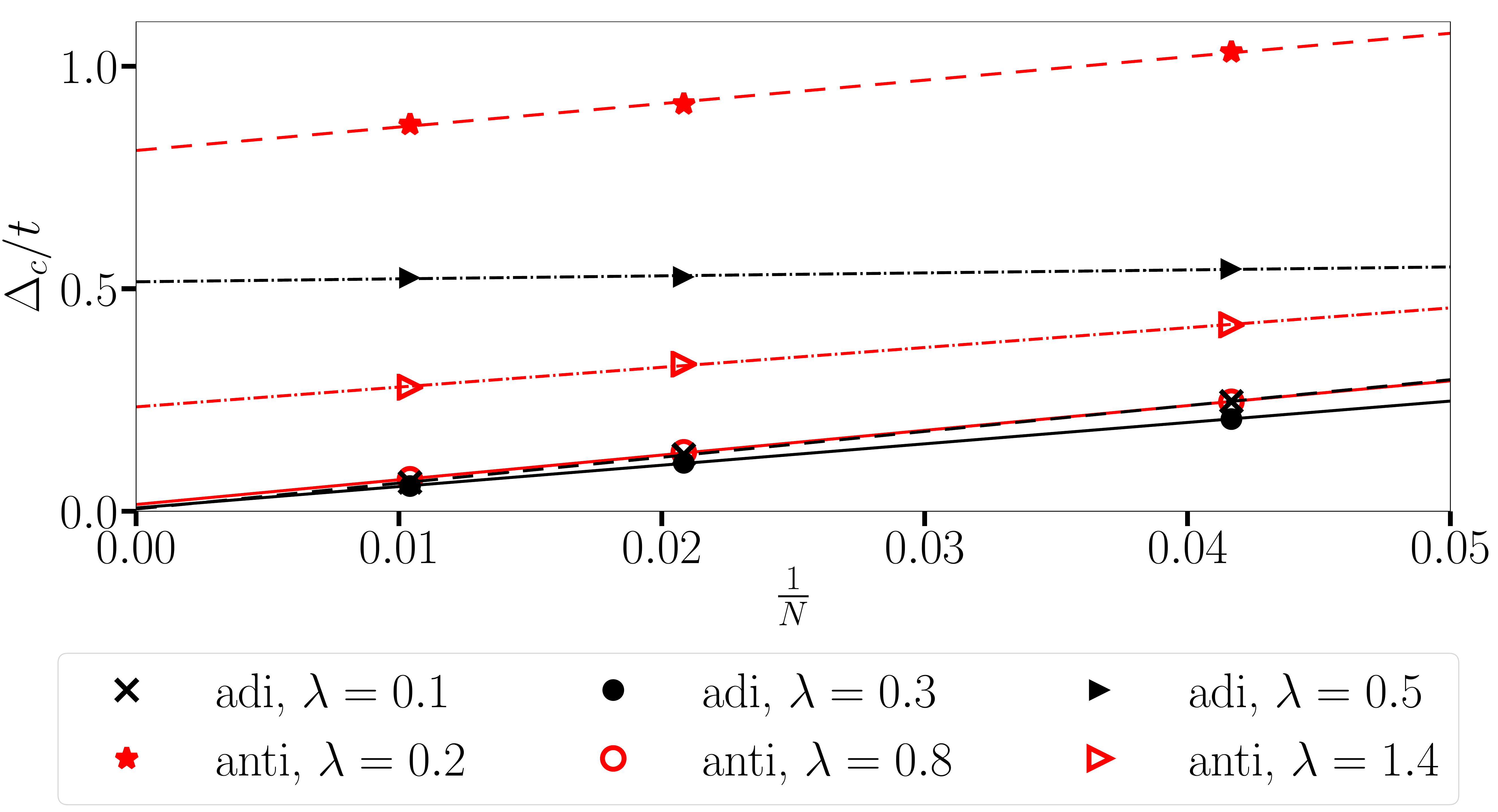}
    \caption{Finite size scaling for the energy gap in the DMRG calculation. We show  some examples, both for the adiabatic limit $(\lambda=0.1;0.3;0.5$ and $u=0.2)$ as well as for the anti-adiabatic limit $(\lambda=0.2;0.8;1.4$ and $u=1.0)$. The extrapolation is done with system sizes of $N=24;48;96$. The scaling is again linear, making it possible remove finite size effects.}
    \label{graph:finite_size_dmrg_gap}
\end{figure}

The Hubbard-Holstein model is defined in infinite space and translationally invariant. Numerically though, we are only able to consider finite systems and therefore have to consider finite size effects and the influence of the boundaries on the observables. This is why, both for the DMRG as well as for the DMET calculation, we do a finite size scaling.\\
In the DMET method, there are two scales to be considered: the size of the whole considered system as well as the impurity size. As the size of the whole system only grows quadratically, we can regard very big systems and therefore did the finite size scaling with system sizes of $N=204$, $N=408$ and $N=816$ sites, as shown in Figure \ref{graph:finite_size_dmet_gap}.
As the other observables, namely the density difference of the electrons between neighboring sites $\left<n_i\right>-\left<n_{i+1}\right>$ and the double occupancy $\left<n_{i\uparrow}n_{i\downarrow}\right>$ are local properties, the finite size effects and the influence of the boundaries do not influence the results anymore for system sizes bigger than $N = 204$.\\
The finite size effects due to the size of the impurity cannot be taken into account that easily, as their scaling is not linear and therefore cannot be rescaled easily. We show the scaling for the energy gap and different impurity sizes in Figure \ref{graph:finite_size_imp} for adiabatic limit (left hand side) and the anti-adiabatic limit (right hand side). With growing impurity sites, the estimation of the energy gap in the Peierls phase gets smaller for both cases and the discrepancy between the results for growing impurity sites get smaller. The scaling is not linear though, making it hard to give a quantitative error estimate.
In the DMRG calculation, opposed to the DMET calculation, we only have two sources of possible errors due to finite size effects, that are the system size itself and the maximal number of considered basis functions in the phononic Fock space, $N_{\rm phon}$. The numerical costs of these calculation also grow polynomially with growing system sizes. This is why, for our extrapolation, we chose to consider system sizes of $N=24$, $N=48$ and $N=96$, as can be seen in Figure \ref{graph:finite_size_dmrg_gap}

\subsection{Energies}\label{section:energies}
\begin{figure}[t]
	\includegraphics[width=.5\textwidth]{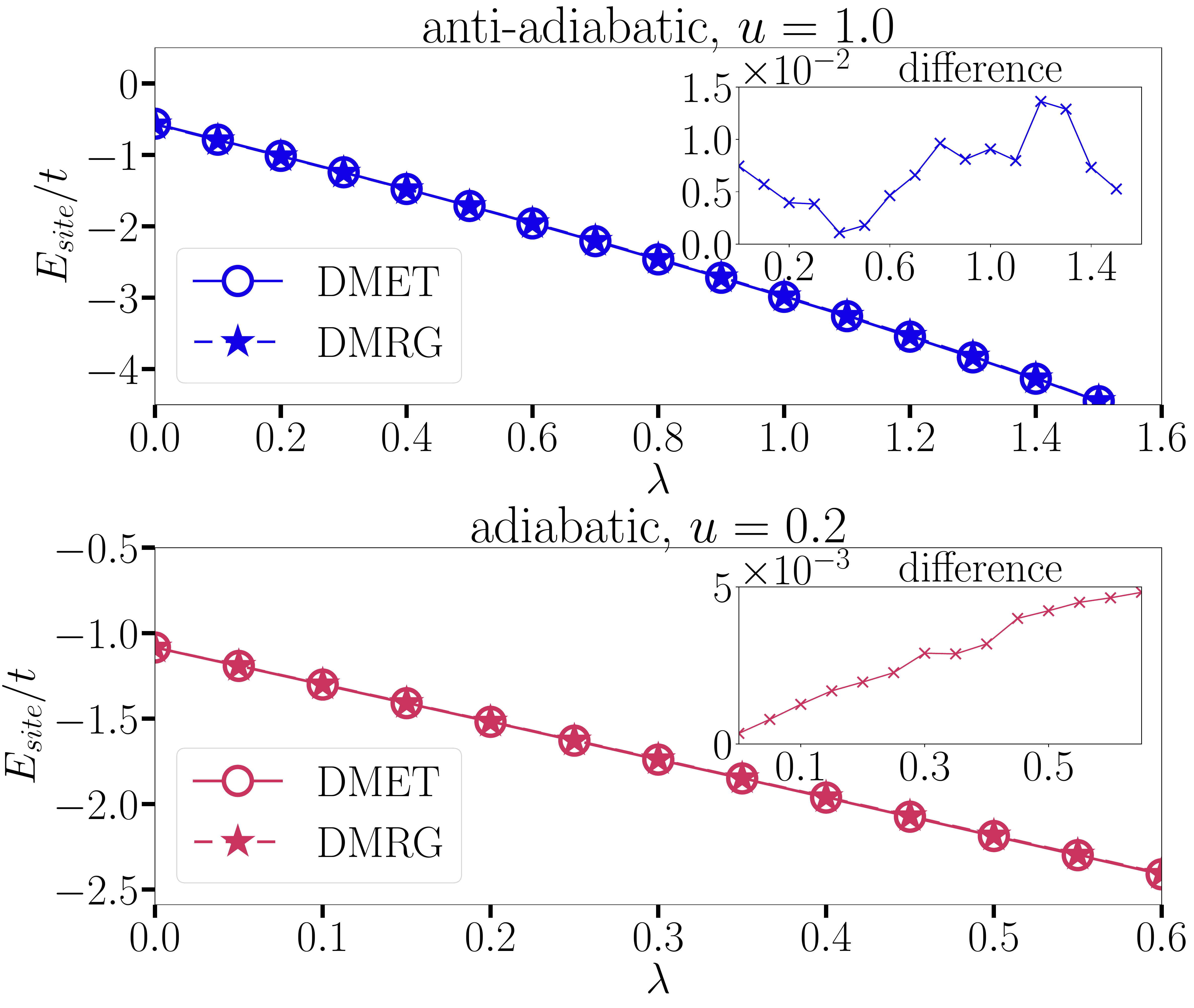}
    \caption{Comparison of the energy per site $E_{\rm site}$, calculated with the DMRG and with the DMET method. In the upper graph, we show the anti-adiabatic limit $(\alpha = 5.0,\,\,\,u=1.0)$, in the lower graph, the adiabatic limit $(\alpha = 0.5,\,\,\,u=0.2)$. For both limits, the results agree quantitatively.}
    \label{graph:energy_dmrg_dmet}
\end{figure}

\begin{figure}[t]
	\includegraphics[width=.5\textwidth]{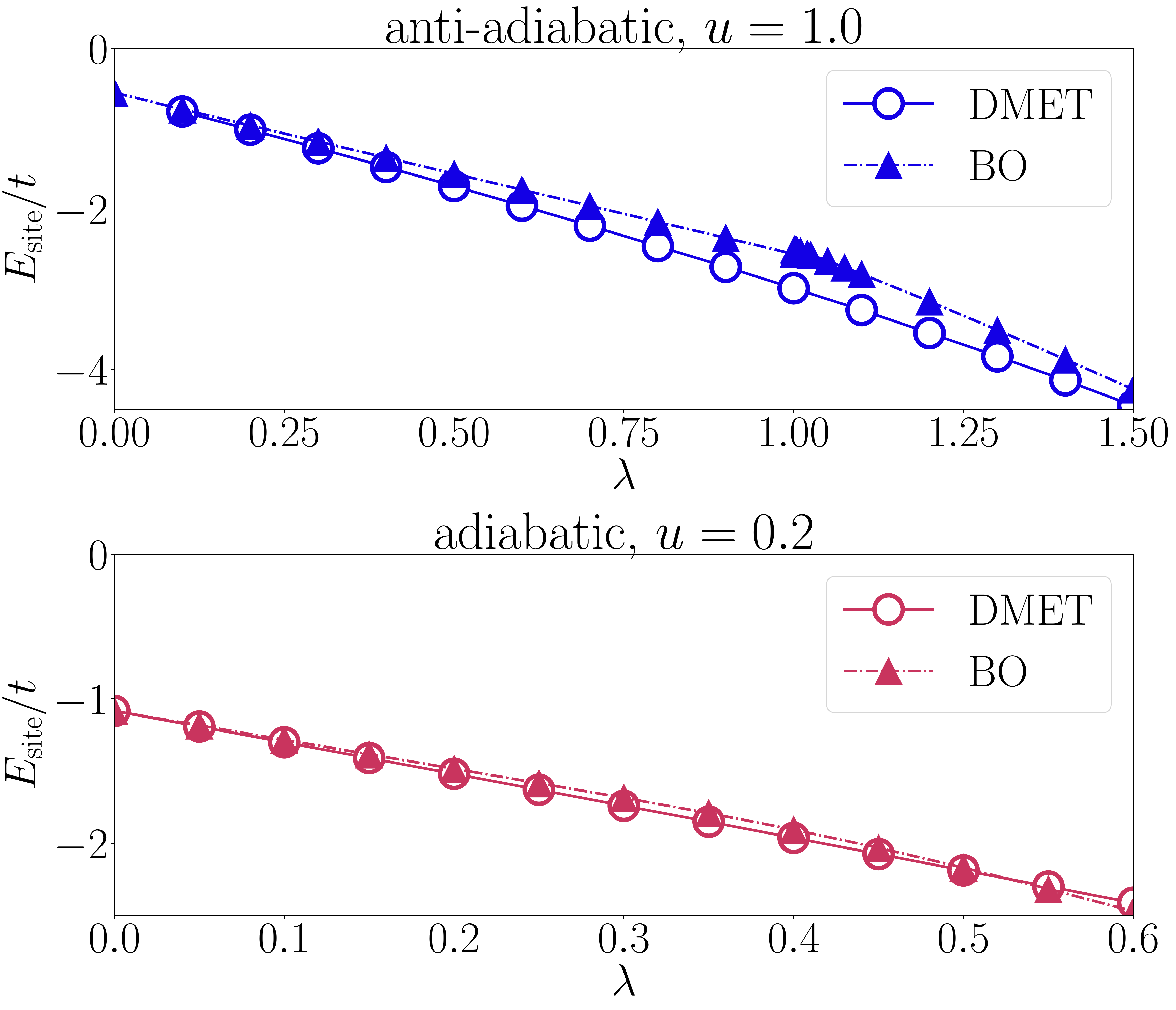}
    \caption{Comparison of the energy per site $E_{\rm site}$ for the full electron-phonon system (DMET calculation), with the energy per site from the same system under the BO approximation. In the upper graph, we show the anti-adiabatic $(\alpha = 5.0,\,\,\,u=1.0)$, in the lower graph the adiabatic limit $(\alpha = 0.5,\,\,\,u=0.2)$. While the behavior only approximately coincides for the anti-adiabatic case, a qualitative agreement between the two methods for the adiabatic limit can be observed.}
    \label{graph:energy_bo_dmet}
\end{figure}

Although not being an observable of physical interest, the energy per site is an important property to show how well two methods agree with each other.\\
In order to benchmark the results of the DMET calculation, we therefore compare the results for the calculated energy per site $E_{\rm site}$ with those from the DMRG calculation. In Figure \ref{graph:energy_dmrg_dmet}, we show the energy per site for the anti-adiabatic $(\alpha = 5.0,\,\,\,u=1.0)$ as well as for the adiabatic limit $(\alpha = 0.5,\,\,\,u=0.2)$ for DMRG and DMET calculations.
For both cases, the results agree on a quantitative level.\\
Additionally, we also compare the energies per site between the full Hubbard-Holstein model and the Hubbard-Holstein model with BO approximation in figure \ref{graph:energy_bo_dmet}. For the anti-adiabatic limit, the energy per site shows approximately the same  behavior while not agreeing quantitatively. In the adiabatic limit, a qualitative agreement can be observed.
\end{document}